\documentclass[12pt,epsf]{article} 

\usepackage{amsmath,amssymb,bm}
\usepackage{amsthm}
\usepackage{mathrsfs}
\usepackage[dvipdfmx]{graphicx}
\usepackage{color}
\usepackage{cancel}

\numberwithin{equation}{section}
\newcommand{\bulkg}{G}
\newcommand{\bulkD}{\nabla}
\newcommand{\bulkR}{R}
\newcommand{\bdyg}{\mathcal{G}}
\newcommand{\bdyD}{\mathcal{D}}
\newcommand{\bdyR}{\mathcal{R}}
\newcommand{\pbdyg}{\delta \bdyg}
\newcommand{\indg}{g}
\newcommand{\indD}{D}
\newcommand{\indR}{R}
\newcommand{\indK}{K}
\newcommand{\indcg}{\Bar{g}}

\newcommand{\indcK}{\Bar{K}}
\newcommand{\pExp}[1]{\langle~#1~\rangle}
\newcommand{\calA}{\mathcal{A}}
\newcommand{\calF}{\mathcal{F}}
\newcommand{\calJ}{\mathcal{J}}
\newcommand{\calK}{\mathcal{K}}

\newcommand{\calS}{\mathcal{S}}
\newcommand{\calT}{\mathcal{T}}
\newcommand{\calZ}{\mathcal{Z}}
\newcommand{\tilA}{\Tilde{A}}
\newcommand{\tcA}{\Tilde{\calA}}
\newcommand{\tos}{\text{on-shell}}
 \newcommand{\tbdy}{{\Sigma_0}}
\newcommand{\tbulk}{\text{bulk}}
\newcommand{\tct}{\text{ct}}
\newcommand{\teff}{\text{eff}}
\newcommand{\tBH}{\text{BH}}
\newcommand{\tExt}{\text{ext}}
\newcommand{\tEin}{\text{Ein}}

\newcommand{\mfw}{\mathfrak{w}}
\newcommand{\mfq}{\mathfrak{q}}
\newcommand{\odiff}[2]{ \frac{d #1}{d #2} }
\newcommand{\odiffII}[2]{ \frac{d^2 #1}{d #2^2} }
\newcommand{\pdiff}[2]{ \frac{\partial #1}{\partial #2} }

\bmdefine{\bms}{\bm{s}}

\makeatletter
\newcommand{\subsubsubsection}{\@startsection{paragraph}{4}{\z@}%
  {1.0\Cvs \@plus.5\Cdp \@minus.2\Cdp}%
  {.1\Cvs \@plus.3\Cdp}%
  {\reset@font\sffamily\normalsize}
}
\makeatother
\begin{document} 

\begin{center}
{\fontsize{20}{0pt}\selectfont Semiclassical Einstein equations from holography and boundary dynamics} 
\end{center}

\medskip
\renewcommand{\thefootnote}{\fnsymbol{footnote}}

\begin{center}
{\large Akihiro Ishibashi}${}^{1}$\footnote{e-mail: akihiro at phys.kindai.ac.jp}, 
{\large Kengo Maeda}${}^{2}$\footnote{e-mail: maeda302 at sic.shibaura-it.ac.jp}, 
{\large and Takashi Okamura}${}^{3}$\footnote{e-mail: tokamura at kwansei.ac.jp}

\medskip

\emph{\small ${}^1$Department of Physics and Research Institute for Science and Technology, \\
Kindai University, Higashi-Osaka, Osaka 577-8502, Japan
\\
\medskip
${}^2$Faculty of Engineering, Shibaura Institute of Technology, \\ 
Saitama 330-8570, Japan
\\
and
\\
${}^3$Department of Physics and Astronomy, Kwansei Gakuin University, \\ 
Sanda, Hyogo, 669-1330, Japan
}
\end{center}
\medskip

\begin{center}
{\bf Abstract}
\end{center}


\begin{center}
\begin{minipage}{16cm}\small 
In this paper, we consider how to formulate semiclassical problems in the context of the AdS/CFT correspondence, based on the proposal of Compere and Marolf. 
Our prescription involves the effective action with self-action term for boundary dynamical fields, which can be viewed as imposing mixed boundary conditions for the gravity dual. We derive the semiclassical Einstein equations sourced by boundary CFT stress-energy tensor. Analyzing perturbations of the holographic semiclassical Einstein equations, we find a universal parameter $\gamma_d$ which controls the contribution from boundary CFTs and specifies dynamics on the AdS boundary. As a simple example, we examine the semiclassical Einstein equations in $3$-dimensions with $4$-dimensional AdS gravity dual, and show that the boundary BTZ black hole with vanishing expectation value of the stress-energy tensor becomes 
unstable due to the backreaction from quantum stress-energy tensor when the parameter $\gamma_d$ exceeds a certain critical value. 
\end{minipage} 

\end{center} 

\renewcommand{\thefootnote}{\arabic{footnote}}

\section{Introduction} 
One of the striking features of the original AdS/CFT correspondence~\cite{Maldacena:1997re,Gubser:1998bc,Witten:1998qj} is that a theory of gravitation is mapped into a theory of entirely non-gravitational quantum fields. This mapping is formulated by imposing the Dirichlet boundary condition for the bulk metric and by taking the boundary metric as a non-dynamical source term for boundary quantum fields. Afterward, the role of boundary conditions in AdS/CFT correspondence has been studied in more detail and it turned out that modifications of the boundary conditions correspond to various deformations of boundary CFTs. A number of possible deformations of AdS/CFT correspondence have been considered, such as the inclusion of multi-trace interactions and some dynamical fields on the boundary~\cite{Witten:2001ua}. An interesting study along this line is the proposal by Compere and Marolf~\cite{Compere:2008us}, in which the boundary metric is promoted to a dynamical field induced by the boundary CFTs. This is done by adding certain boundary counter-terms to the action, the prescription of which essentially corresponds to changing the boundary conditions for the bulk gravity and has been applied to, for instance, cosmology in semiclassical regime \cite{Ecker:2021cvz}, as well as holographic superconductors~\cite{Natsuume:2022kic}
\footnote{In~\cite{Natsuume:2022kic}, instead of gravity, boundary Maxwell fields are promoted to dynamical fields satisfying semiclassical Maxwell equation coupled to boundary $U(1)$ current. See also \cite{Ahn:2022azl} for boundary dynamical gauge fields with mixed boundary conditions, and \cite{Ecker:2018ucc} for an application to heavy ion physics.  
}. 
It is of considerable interest to further study dynamical gravity coupled with quantum fields in the context of deformed AdS/CFT correspondence.

The main purpose of this paper is to present a general scheme to formulate the semiclassical Einstein equations on the AdS conformal boundary, following the idea of \cite{Compere:2008us}, and discuss possible uses of such holographic semiclassical equations, providing some simple example. In order to treat the boundary metric as a dynamical variable that satisfies the Einstein equations with the source of vacuum expectation value for strongly interacting CFT stress-energy tensor, the bulk metric is required to satisfy general (mixed-)boundary conditions, rather than the Dirichlet boundary condition at the AdS boundary. Changing boundary conditions can be achieved by adding the Einstein-Hilbert term to the boundary CFT action. 
In other words, the semiclassical Einstein equations for the boundary metric plays the role of mixed-boundary conditions 
for the bulk Einstein equations. 
We explicitly describe our prescription for obtaining holographic semiclassical Einstein equations in the case that the bulk and boundary geometries are given by the $d+1$ and $d$-dimensional AdS metrics with boundary CFTs being in the conformal vacuum state. 
By inspecting bulk and boundary perturbations, we find that the following dimensionless parameter [see eq.~(\ref{eq:def-Upsilon:in:d}) below] controls the contribution from $d$-dimensional boundary CFT:  
\begin{equation}
 \gamma_d := \dfrac{G_d L}{\pi G_{d+1}}\left( \dfrac{L}{\ell} \right)^{d-2}   \,,
\label{def:param} 
\end{equation} 
where $G_{d+1}, G_d$, and $L$, $\ell$, denote the bulk and boundary gravitational couplings and corresponding curvature radius, respectively.  
This is the universal parameter, common to all our geometrical settings, that represents the ratio of the strength of the boundary quantum stress-energy $\langle {\cal T}_{\mu \nu} \rangle$ with respect to that of the boundary cosmological constant $\Lambda_d$, 
and that specifies boundary dynamics, e.g., stability property and phase transition. 
%
%
As a concrete example of the application of our scheme, we study linear perturbations of the $4$-dimensional AdS bulk with the BTZ boundary metric and demonstrate that the BTZ black hole with vanishing expectation value of the stress-energy tensor becomes unstable due to the backreaction of boundary quantum stress-energy, when the parameter (\ref{def:param}) exceeds a certain critical value.

This paper is organized as follows. In the next section, we will provide a general prescription for constructing semiclassical problems including 
the derivation of semiclassical Maxwell equations coupled with R-current, as well as semiclassical Einstein equations coupled with boundary CFTs. To be concrete we in particular apply the prescription to derive the semiclassical Einstein equations in $d$-dimensional AdS boundary and study perturbations thereof. We find the universal parameter $\gamma_d$. Then, in section~\ref{sec:3}, we consider the case in which the boundary geometry is 
described by BTZ metric and show that due to the backreaction from boundary CFT stress-energy, BTZ black hole becomes unstable when our 
universal parameter exceeds the critical value. Section~\ref{sec:4} is devoted to summary and discussions. 
As an application of our holographic semiclassical formulations, we show how to calculate permittivity and permeability by using the holographic semiclassical Maxwell theory at finite temperature in Appendix~\ref{sec:Maxwell_exmpl}. 
In Appendix~\ref{sec:calT}, we provide concrete expression of the vacuum expectation value of CFT stress-energy tensor. Appendix~\ref{sec:formula-holo-semiC_Ein_eq} provides some detailed formulas for perturbations of the holographic semiclassical Einstein equations. 

\section{Holographic semiclassical problems}
\label{sec:formalism}
In this section, we describe how to formulate semiclassical problems in the  context of holography. 
Our main interest is in the boundary Einstein gravity sourced by the expectation values of CFT's stress-energy tensor, 
but the prescription works for a broader class of semiclassical problems. 

\subsection{Notation and conventions}
\begin{figure}[htbp]
  \begin{center}
  \includegraphics[width=9cm]{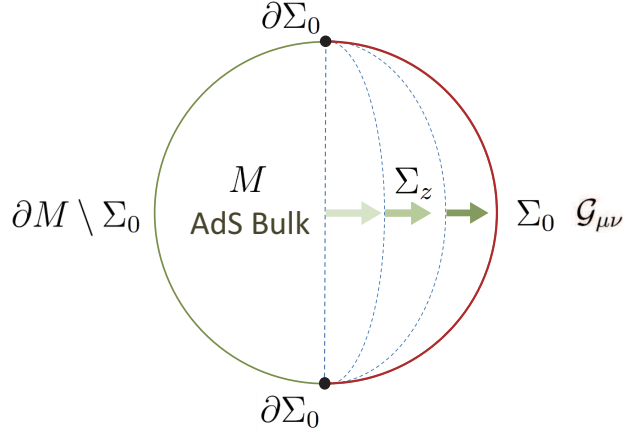}
  \end{center}
 \vspace{-0.5cm} 
  \caption{
In the $(d+1)$-dimensional bulk $M$, we consider a foliation by $d$-dimensional hypersurfaces $\{\Sigma_z\}$, which has the limit $\Sigma_0$ 
on the AdS conformal boundary $\partial M$.}
 \label{geom}
\end{figure}

We are concerned with $(d+1)$-dimensional AdS bulk spacetime $(M,\bulkg_{MN})$ and its conformal completion 
with $d$-dimensional conformal boundary $\partial M$. 
We consider, inside the bulk, a set of $d$-dimensional hypersurfaces $\Sigma_z$, parametrized by $z \in {\Bbb R}^+$ with induced metrics $g_{\mu \nu}$  which foliate (a part of) $M$ and admit the limit hypersurface $ \Sigma_0 := \lim_{z\rightarrow 0} \Sigma_z$ on (a part of) the conformal boundary 
$\partial M$. We also introduce, on each $\Sigma_z$, a $d$-dimensional metric $\tilde{g}_{\mu \nu} \:= \Omega^2 g_{\mu \nu}$ with an appropriate function $\Omega$ so that the limit ${\cal G}_{\mu \nu}:=\lim_{z \to 0} \tilde{g}_{\mu \nu}$ provides a regular metric on $\Sigma_0 \subset \partial M$. See figure~\ref{geom}. 
Our primal interest is this metric ${\cal G}_{\mu \nu}$, which is promoted to be a boundary dynamical field. 
As already done above, we express tensors in $(d+1)$-dimensions with upper case latin indices $M, N, \dots$ and those in $d$-dimensional hypersurfaces with greek indices. So, for example, a vector field in the bulk is denoted by $A_M$, whose pull-back to $\Sigma_z$ is by $A_\mu$, whereas the corresponding vector field on $\Sigma_0 \subset \partial M$ by $\calA_\mu$. 
When we need to distinguish which metrics, either ${\cal G}_{\mu \nu}$, $g_{\mu \nu}$, or $\tilde{g}_{\mu \nu}$, tensors under consideration are associated with, we will explicitly write them as the argument so that for instance, the curvature tensor on the conformal boundary is 
expressed by ${\cal R}_{\mu \nu \rho \sigma} = R_{\mu \nu \rho \sigma}[{\cal G}]$. 

In a neighborhood of given $\Sigma_z$ one can always take, at least locally, a coordinate system $X^M=(z, x^\mu)$ in which 
\begin{eqnarray}
 ds_{d+1}^2 &=&  G_{MN}\, dX^M dX^N
\nonumber \\ 
  &=& \Omega^{-2}(z)\, dz^2 + \indg_{\mu\nu}(x, z)\, dx^\mu dx^\nu 
\nonumber \\
  &
  =& \Omega^{-2}(z)\, \left\{ dz^2 + \tilde{g}_{\mu\nu}(x, z)\, dx^\mu dx^\nu \right\}
  ~. 
\label{eq:bulk_metric-cano_form}
\end{eqnarray} 
In this coordinate system, the metric on the conformal boundary is 
\begin{align}
  & \bdyg_{\mu\nu}(x)
  = \lim_{z \to 0} \Omega^2(z) { \indg_{\mu\nu}(x, z)}
  = \lim_{z \to 0} \left( \frac{z}{L} \right)^2\, \indg_{\mu\nu}(x, z)
  ~, 
\label{eq:bulk_metric-fall_off}
\end{align} and the extrinsic curvature $\indK_{\mu\nu}$ of $\Sigma_z$ is defined by
\begin{align}
  & \indK_{\mu\nu}
  = - \frac{1}{2} \Omega {\partial_z g_{\mu\nu} }
  ~. 
\label{eq:def-ext_curvature}
\end{align}

It is sometime more convenient to use the conformally rescaled metric, $\tilde{g}_{\mu\nu}  = \Omega^2\, \indg_{\mu\nu}$ 
with the conformal factor behaving as $\Omega(z) \xrightarrow{z \to 0} {z}/{L}$, and the corresponding extrinsic curvature, 
\begin{align}
  & \tilde{K}_{\mu\nu}
  := - \frac{1}{2}\, \partial_z \, \tilde{g}_{\mu\nu}
  = \Omega\, \indK_{\mu\nu} -  \indg_{\mu\nu}\, \Omega\, \Omega'(z)
  ~. 
\label{eq:def-barK}
\end{align}
%
Here and hereafter the {\em prime} \lq\lq\, ${~}'$\, \rq\rq\ denotes the derivative by $z$, and 
the indices of tensors with \lq\lq\, $\tilde{~~}$\, \rq\rq\ are raised and lowered by $\tilde{g}_{\mu\nu}$, $\tilde{g}^{\mu\nu}$. 
For later convenience, we summarize our notation and conventions in Table 1.   
\begin{table}[h]
\centering
\caption[A summary of notation]{A summary of notation}
\begin{tabular}{l|cccc}
        & metric \& index & covariant derivative & curvature & length scale \\ \hline
      bulk $M$ & $\bulkg_{MN}$ & $\bulkD_M$ & $\bulkR_{KLMN}[\, \bulkg\, ]$ & $L$
  \\
   boundary $\Sigma_0 \subset \partial M$ & $\bdyg_{\mu\nu}$
     & $\bdyD_\mu$
     & $\bdyR_{\mu\nu\rho\sigma} := R_{\mu\nu\rho\sigma}[\, \bdyg\, ]$ & $\ell$
  \\
   hypersurface $\Sigma_z$ & $\indg_{\mu\nu}$ & $\indD_\mu$
     & $R_{\mu\nu\rho\sigma}[\, \indg\, ]$
  \\
   hypersurface $\Sigma_z$ & $\tilde{g}_{\mu\nu}$ & $\tilde{D}_\mu$
     & $\tilde{R}_{\mu\nu\rho\sigma} := R_{\mu\nu\rho\sigma}[\, \tilde{g}\, ]$
\end{tabular}
\end{table}
%

 
\subsection{Holographic semiclassical Maxwell equations}
\label{sec:U1_exmpl}

Before going into the semiclassical Einstein equations, to illustrate our basic idea in a simpler example, we first consider the semiclassical problem of a $3$-dimensional U(1) gauge field $\calA_\mu$ coupled with R-current $\pExp{\calJ^\mu}$, studied in \cite{Natsuume:2022kic}\footnote{
Note that although R-symmetry is a global symmetry, it is promoted to a local symmetry when it couples with $\calA_\mu$~\cite{Maeda:2010br}.}. 
Our starting point is the effective action $\calS_\teff$, obtained from the partition function $\calZ[\, \calA\, ]$ of the boundary gauge field $\calA_\mu$ as
\begin{align}
  & \calS_\teff
  = - \frac{1}{4\, e^2}\, \int d^3x~\calF^{\mu\nu} \calF_{\mu\nu}
  - i \ln\calZ[\, \calA\, ]
  ~,
\label{eq:q_action-Maxwell_exmpl}
\end{align}
where $e$ denotes the U(1) coupling  constant. The partition function gives rise to the expectation value of the current as  
\begin{equation}
\pExp{\calJ^\mu}  := - i\, \dfrac{\delta \ln\calZ[\, \calA\, ] }{\delta \calA_\mu} \,.
\end{equation} 
Applying variational principle to $\calS_\teff$, we obtain the semiclassical Maxwell equations for the gauge 
field, 
\begin{align}
  & \frac{1}{e^2}\, \partial_\nu \calF^{\mu\nu}
  = \pExp{\calJ^\mu}
  ~. 
\label{eq:q_EOM-Maxwell_exmpl}
\end{align}
It is in general difficult to evaluate $\calZ[\, \calA\, ]$ as a functional of $\calA_\mu$, but the AdS/CFT correspondence makes it possible  
to compute $\calZ[\, \calA\, ]$ in terms of the classical solutions of the bulk U(1) gauge field $A_M$ in the asymptotically AdS$_{4}$ as follows:
\footnote{Note that for R-current in AdS$_{4}/$CFT$_3$, counter terms are not needed. }
\begin{subequations}
\label{eq:partition_function-Maxwell_exmpl-all}
\begin{align}
  & - i\, \ln\calZ[\, \calA\, ]
  = S^{\tos}_\tbulk[\, \calA\, ]
  ~,
\label{eq:partition_function-Maxwell_exmpl} \\
  & S_\tbulk[\, A\, ]
  = - \frac{1}{ 4\, g^2 }\,
  \int_{M} d^{4}X~\sqrt{- \bulkg}~F^{MN} F_{MN}
  ~, 
\label{eq:bulk_action-Maxwell_exmpl}
\end{align}
\end{subequations}
where the classical solution $A_M$ is supposed to satisfy the Dirichlet boundary condition $A_\mu \big|_{\tbdy} = \calA_\mu$ 
at the boundary $\Sigma_0$\footnote{$A_M$ should also satisfy certain boundary or regularity conditions inside the bulk.}.

In general, when the source term for some field of interest has a gravitational dual in the holographic context, 
one can expect to formulate the semiclassical problem of the field by adding the self-action and by applying the variational principle.  
To see this is indeed the case, let us consider, instead of (\ref{eq:q_action-Maxwell_exmpl}), 
the following effective action, 
\begin{align}
  & \calS_\teff
  = \int d^3x~\left( - \frac{1}{4\, e^2}\, \calF^{\mu\nu} \calF_{\mu\nu}
      + \calJ_\tExt^\mu \calA_\mu \right)
    + S^{\tos}_\tbulk[\, \calA\, ]
\nonumber \\
  &\hspace*{0.6truecm}
  = \int_\tbdy d^3x~\left( - \frac{1}{4\, e^2}\, \calF^{\mu\nu} \calF_{\mu\nu}
      + \calJ_\tExt^\mu \calA_\mu \right)
  - \frac{1}{ 4\, g^2 }\,
    \int_{M} d^{4}X~\sqrt{- \bulkg}~F^{MN} F_{MN}
  ~.
\label{eq:holo_semiC-action-Maxwell_exmpl}
\end{align}
Note that for later convenience, we have added the extra term $\calJ_\tExt^\mu$ which describes all contributions to the current other than that from the 
dual field theory. Taking the variation of this action will provide the desired semiclassical equations of motion for $\calA_\mu$. In the variation, one does 
not need to impose the Dirichlet condition at the AdS boundary. Instead, the semiclassical equations of motion take the place of the boundary conditions.  

Let us see, in more detail, the above formulation of the semiclassical problem~(\ref{eq:holo_semiC-action-Maxwell_exmpl}) 
in a concrete setting. Consider a $4$-dimensional asymptotically AdS bulk geometry, whose metric takes the asymptotic form in the Poincar\'e chart 
$X^M = (z, x^\mu), \: x^\mu = (t, x, y)$ as 
\begin{align}
  & ds_{4}^2
  \xrightarrow{z \to 0} \left( \frac{L}{z} \right)^2\,
    \left( dz^2 + \eta_{\mu\nu}\, dx^\mu dx^\nu \right)
  ~.
\label{eq:asympt-bulk_metric}
\end{align}
With variations of $A_M$ and $\calA_\mu$, the action (\ref{eq:holo_semiC-action-Maxwell_exmpl}) changes as
%
\begin{align}
  & \delta\calS_\teff
  = \frac{1}{e^2}\, \int_\tbdy d^3x~\partial_\nu
    \big( \calF^{\mu\nu}\, \delta\calA_{\mu} \big)
  + \int_\tbdy d^3x\left( - \frac{ \partial_\nu \calF^{\mu\nu} }{e^2}
    + \calJ_\tExt^\mu + \frac{ \eta^{\mu\nu} }{ g^2 }\, F_{z\nu}
  \right)\, \delta \calA_{\mu}
\nonumber \\
  &\hspace*{1.5truecm}
  + \frac{1}{2\, g^2}\, \int_{M} d^{4}X~\sqrt{ - \bulkg }~
    \big( \bulkD_M F^{MN} \big)\,
    \delta A_{N}
  ~, 
\label{eq:delta_S-holo_semiC-Maxwell_exmpl-2}
\end{align}
%
where the contribution to $\delta S^{\tos}_\tbulk$ is assumed to stem only 
from $\Sigma_0$
\footnote{This is equivalent to assuming that $S^{\tos}_\tbulk$ be a functional only of the boundary field $\calA_\mu$.}. 
Since, according to the AdS/CFT correspondence, the expectation value of the R-current is given by 
%
\begin{align}
  & \pExp{\calJ^\mu}
  := \frac{\delta}{ \delta \calA_\mu }\,
    S^{\tos}_\tbulk[\, \calA\, ]
  = \frac{ \eta^{\mu\nu} }{ g^2 }\, F_{z\nu}\, \big|_{z=0}
  ~,
\label{eq:dictionary-current}
\end{align}
when $\bulkD_N F^{MN} = 0$ holds in the bulk, eq.~(\ref{eq:delta_S-holo_semiC-Maxwell_exmpl-2}) becomes 
\begin{align}
  & \delta\calS_\teff
  = \frac{1}{e^2}\, \int_\tbdy d^3x~\partial_\nu
    \big( \calF^{\mu\nu}\, \delta\calA_{\mu} \big)
  + \int_\tbdy d^3x\left( - \frac{ \partial_\nu \calF^{\mu\nu} }{e^2}
    + \calJ_\tExt^\mu + \pExp{\calJ^{\mu}} \right)\, \delta \calA_{\mu}
  ~.
\label{eq:delta_S-holo_semiC-Maxwell_exmpl-3}
\end{align}
In this way, applying the variational principle to eq.~(\ref{eq:holo_semiC-action-Maxwell_exmpl}) and noting $\delta \calA_\mu \ne 0$, 
we obtain the bulk equations of motion $\bulkD_N F^{MN} = 0$ and the boundary 
semiclassical Maxwell equations~(\ref{eq:q_EOM-Maxwell_exmpl}). 
As a consequence, the semiclassical problem (\ref{eq:q_EOM-Maxwell_exmpl}) in the holographic context reduces to the problem of 
solving the following set of the bulk classical Maxwell equations and the boundary semiclassical Maxwell equations,   
\begin{subequations}
\label{eq:EOM-holo_semiC-Maxwell_exmpl-all}
\begin{align}
  & 0
  = \bulkD_N F^{MN}
  ~,
\label{eq:EOM-holo_semiC-Maxwell_exmpl-bulk} \\
  & 0
  = \partial_\nu \calF^{\mu\nu}
  - \frac{e^2}{g^2}\, \eta^{\mu\nu} F_{z\nu}\, \big|_{z=0}
  - e^2\, \calJ_\tExt^\mu
  ~, 
\label{eq:EOM-holo_semiC-Maxwell_exmpl-bdy}
\end{align}
\end{subequations}
under appropriate boundary conditions, if needed, in the bulk. Note that setting $e^2 \rightarrow 0$ corresponds to the Dirichlet boundary condition, since in this limit, 
the (finite) self-action for $\cal{A}_\mu$ decouples from the bulk action in (\ref{eq:holo_semiC-action-Maxwell_exmpl}), and thus can be regarded as a background field. 
On the other hand, the limit $e^2 \rightarrow \infty$ corresponds to the Neumann conditions~\cite{Domenech:2010nf} (see also \cite{Montull:2009fe}), since in this case, with setting ${\cal J}_{\rm ext}=0$, one finds 
$F_{z \mu}|_{\rm bdy}=0$.
A simple application of the above set of equations is given in Appendix~\ref{sec:Maxwell_exmpl}.


\subsection{Holographic semiclassical Einstein equations} 

The prescription for the holographic semiclassical problem illustrated in Sec.~\ref{sec:U1_exmpl} applies to the derivation of semiclassical Einstein equations. Our effective action for boundary semiclassical problem is now the following: 
\begin{align}
  & \calS_\teff
  = \int_\tbdy \frac{ d^dx\, \sqrt{- \bdyg} }{ 16\, \pi\, G_{d} }\,
    \left( \bdyR - 2\, \Lambda_{d} \right)
    + S_\tbulk[\, \bulkg\, ]
  ~, 
\label{eq:holo_semiC-action-Einstein_eqn}
\end{align}
with $G_d$, $\Lambda_d$ being the $d$-dimensional gravitational constant and cosmological constant, respectively.
Here $S_\tbulk$ presents the action for the gravity dual used in the standard AdS/CFT correspondence\footnote{As can be seen in eqs.~(\ref{eq:bulk_action-Einstein_eqn-all}), in this paper, by ``bulk action" $S_\tbulk$ we mean the action which includes 
counter-terms $S_\tct$ so that $S_\tbulk$ is already regularized and therefore its on-shell value is a functional of the boundary dual ${\cal G}_{\mu \nu}$.}. For purely gravitational case, our bulk action $S_\tbulk$ is given by 
\begin{subequations}
\label{eq:bulk_action-Einstein_eqn-all}
\begin{align}
  & S_\tbulk[\, \bulkg\, ]
  = S_{\text{EH}+\text{GH}}[\, \bulkg\, ] + S_\tct[\, \indg\, ]
  ~,
\label{eq:bulk_action-Einstein_eqn} \\
  & S_{\text{EH}+\text{GH}}[\, \bulkg\, ]
  = \int_{M} \frac{ d^{d+1}X\, \sqrt{- \bulkg} }{ 16\, \pi\, G_{d+1} }\,
    \left\{ \bulkR[\, \bulkg\, ] + \frac{d\, (d - 1)}{L^2} \right\}
  + \int_{\tbdy} \frac{ d^{d}x\, \sqrt{-\indg} }{ 8\, \pi\, G_{d+1} }\, \indK
  ~,
\label{eq:EH+GH_action-Einstein_eqn} \\
  & S_\tct[\, \indg\, ]
  = - \int_{\tbdy} \frac{ d^{d}x\, \sqrt{-\indg} }{ 16\, \pi\, G_{d+1} }\,
  \left\{ 2\, \frac{ d - 1 }{L} + \frac{L}{d - 2}\, \indR[\, \indg\, ]
  + \cdots \right\}
  ~, 
\label{eq:ct_action-Einstein_eqn}
\end{align}
\end{subequations}
where $G_{d+1}$ and $L$ denote, respectively, the $(d+1)$-dimensional gravitational coupling and the curvature length\footnote{
The corresponding cosmological constant is $2\, \Lambda_{d+1} = - d (d - 1)/L^2$.}, and where in $S_\tct$, we have expressed, for simplicity, only terms which are required in $d \le 3$. 
Note that as is often the case, we first evaluate the on-shell value of each term in eq.~(\ref{eq:bulk_action-Einstein_eqn}) on a certain ``cutoff hypersurface" and then take the limit toward the conformal boundary, 
accordingly $\indK$ is also evaluated as the trace of the extrinsic curvature of the cutoff hypersurface and taken the boundary limit. 
Now let us apply the variational principle to $\calS_\teff$ (\ref{eq:holo_semiC-action-Einstein_eqn}).  
When $S_\tbulk$ is given by eq. ~(\ref{eq:bulk_action-Einstein_eqn-all}), the variation of $\calS_\teff$ reduces to 
\begin{align}
  & \delta\calS_\teff
  = \int_\tbdy d^dx~\partial_\mu
    \left( \frac{ \sqrt{- \bdyg} }{ 16\, \pi\, G_{d} }\, \cdots \right)
  - \int_\tbdy \frac{ d^dx\, \sqrt{- \bdyg} }{ 16\, \pi\, G_{d} }\,
    \left( \bdyR^{\mu\nu} - \frac{ \bdyR }{2}\, \bdyg^{\mu\nu}
      + \Lambda_{d}\, \bdyg^{\mu\nu} \right)\, \delta \bdyg_{\mu\nu}
\nonumber \\
  &
  - \int_{M} \frac{ d^{d+1}X\, \sqrt{- \bulkg} }{ 16\, \pi\, G_{d+1} }\,
    \left( \tEin^{MN}[\, \bulkg\, ] + \Lambda_{d+1}\, \bulkg^{MN}
    \right)\, \delta\bulkg_{MN}
  + \int_\tbdy d^dx~
    \frac{ \delta S^{\tos}_\tbulk[\, \bdyg\, ] }{ \delta \bdyg_{\mu\nu} }\,
    \delta \bdyg_{\mu\nu}
  ~
\label{eq:delta-S-holo_semiC-action-Einstein_eqn}
\end{align}
where $\tEin^{MN}$ denotes the Einstein tensor with respect to the bulk metric $G_{MN}$. 
Note that the last term has come from $S_\tbulk$ and where we have considered the boundary term 
contribution only from $\Sigma_0$. 
Now, applying the variational principle with the condition $\delta \bdyg_{\mu \nu} \neq 0$, 
we obtain from (\ref{eq:delta-S-holo_semiC-action-Einstein_eqn})
the following set of the bulk Einstein equations and boundary semiclassical Einstein equations 
\begin{subequations}
\label{eq:EOM-holo_semiC-Einstein_eqn-all}
\begin{align}
  & \tEin^{MN}[\, \bulkg\, ] + \Lambda_{d+1}\, \bulkg^{MN}
  = 0
  ~,
\label{eq:EOM-holo_semiC-Einstein_eqn-bulk} \\
  & \bdyR^{\mu\nu} - \frac{ \bdyR }{2}\, \bdyg^{\mu\nu}
  + \Lambda_{d}\, \bdyg^{\mu\nu}
  = 8\, \pi\, G_{d}\, \pExp{\calT^{\mu\nu}}
  ~, 
\label{eq:EOM-holo_semiC-Einstein_eqn-bdy}
\end{align}
\end{subequations}
where the expectation value of the stress-energy $\pExp{\calT^{\mu\nu}} $ is given in terms of the on-shell value of $S_\tbulk$, by  
\begin{align}
  & \pExp{\calT^{\mu\nu}}
  = \frac{2}{ \sqrt{- \bdyg} }\,
  \frac{ \delta S^{\tos}_\tbulk[\, \bdyg\, ] }{\delta \bdyg_{\mu\nu}}
  ~. 
\label{eq:energy_stress-calZ}
\end{align}
\bigskip 

\noindent 
{\bf Remarks:} 
\begin{itemize}

\item[1)] When our boundary of interest $\Sigma_0$ is a proper subset of $\partial M$, 
we need to impose appropriate boundary conditions (i) at the rest of the conformal boundary $\partial M \setminus \Sigma_0$, 
as well as (ii) at the corner $\partial \Sigma_0$, and (iii) at inner boundaries (e.g., horizons), 
if exist, inside the bulk\footnote{For example, when studying, via eqs.~(\ref{eq:EOM-holo_semiC-Einstein_eqn-all}), boundary dynamical gravity coupled with 
thermal CFTs, one may consider an asymptotically AdS black hole spacetime as the dual bulk. In such a case, one needs to 
impose some regularity condition at the horizon of the bulk black hole, in order to solve eq.~(\ref{eq:EOM-holo_semiC-Einstein_eqn-bulk}).
}. 
One might think of the asymptotic condition~(\ref{eq:bulk_metric-fall_off})
as the imposition of the Dirichlet boundary condition at $\Sigma_0$. However, ${\cal G}_{\mu \nu}$ is now a dynamical variable on $\Sigma_0$ to be determined so as to satisfy eq.~(\ref{eq:EOM-holo_semiC-Einstein_eqn-bdy}). 
Therefore, (i)--(iii) are the places to impose boundary conditions, in order to solve the whole system of eqs.~(\ref{eq:EOM-holo_semiC-Einstein_eqn-all}). We can, in principle, impose the Dirichlet boundary conditions\footnote{Depending upon what kind of holographic setting one wants to consider, it may also be possible 
to impose mixed boundary conditions as a further deformation of the AdS/CFT correspondence. } 
there at (i)--(iii).

\item[2)] Similar to the case of semiclassical problem for Maxwell fields~(\ref{eq:EOM-holo_semiC-Maxwell_exmpl-all}), setting $G_d \rightarrow 0$ corresponds to the Dirichlet boundary condition for the bulk metric, while $G_d \rightarrow \infty$ to the Neumann boundary condition. 

\item[3)] 
One way of solving the boundary value problem~(\ref{eq:EOM-holo_semiC-Einstein_eqn-all}) is to use the shooting method by transforming (\ref{eq:EOM-holo_semiC-Einstein_eqn-all}) into an ``initial value problem" as follows. 
The right-hand side 
of eq.~(\ref{eq:EOM-holo_semiC-Einstein_eqn-bdy}), or (\ref{eq:energy_stress-calZ}), is expressed as  
\begin{subequations}
\label{eq:delta-S_bulk_os-all}
\begin{align}
  & \frac{ \delta S^\tos_\tbulk[\, \bdyg\,] }{\delta \bdyg_{\mu\nu}}
  = \lim_{z \to 0} \frac{L^2}{z^2}\,
    \frac{ \delta S^\tos_\tbulk[\, \indg\, ] }{\delta \indg_{\mu\nu}}
  = 
  \lim_{z \to 0} \frac{L^2}{z^2}\,
  \left( \pi^{\mu\nu} + \frac{\delta S_\tct}{\delta \indg_{\mu\nu}} \right)
  ~,
\label{eq:delta-S_bulk_os} \\
  & \pi^{\mu\nu}
  := \frac{ \delta S^\tos_{\text{EH}+\text{GH}} } {\delta \indg_{\mu\nu}}
  = \frac{ \sqrt{-\indg} }{16\, \pi\, G_{d+1}} \,
    \left( - \indK^{\mu\nu} + \indg^{\mu\nu}\, \indK \right)
  ~, 
\label{eq:def-conjugate_pi}
\end{align}
\end{subequations}
where $\pi^{\mu\nu}$ denotes the momentum canonically conjugate to $\indg_{\mu\nu}$, evaluated on a cutoff-surface via eq.~(\ref{eq:EOM-holo_semiC-Einstein_eqn-bdy}).  
Then, one tries to find solutions of eq.~(\ref{eq:EOM-holo_semiC-Einstein_eqn-all}) as a sort of initial value problem of eq.~(\ref{eq:EOM-holo_semiC-Einstein_eqn-bulk}) by evolving toward inside the bulk from the initial data $(g_{\mu \nu}, \pi^{\mu \nu})$ given (on a cutoff-hypersurface) by eq.~(\ref{eq:EOM-holo_semiC-Einstein_eqn-bdy}), until 
the solution eventually matches the boundary conditions at (i)--(iii).  

\item[4)] 
As we show in Appendix~\ref{sec:calT}, $\pExp{\calT_{\mu\nu}}$ given in (\ref{eq:energy_stress-calZ}) can be expressed 
in terms of the conformally rescaled $\tilde{g}_{\mu\nu}$, $\tilde{K}_{\mu\nu}$---which have the regular boundary limits (see 
eq.~(\ref{eq:def-barK}) for the definition)---as 
%
\begin{align}
  & \pExp{\calT_{\mu\nu}}
  = \lim_{z \to 0} \frac{1}{ 8\, \pi\, G_{d+1}\, L }\,
  \Bigg[\, L^2\,
    \frac{ \tEin_{\mu\nu}[\, \bulkg\, ] + \Lambda_{d+1}\, \bulkg_{\mu\nu} }
         { (d - 2)\, \Omega^{d-2} }
  + L^2\, \frac{ 2\, \tilde{K} \tilde{K}_{\mu\nu}
    - \tilde{g}_{\mu\nu}\, (\tilde{K}^{\rho\sigma} \tilde{K}_{\rho\sigma} + \tilde{K}^2) }
               { 2\, (d - 2)\, \Omega^{d-2} }
\nonumber \\
  &\hspace*{2.0truecm}
  - L\, \tilde{g}_{\nu\rho}\,
    \left( \frac{L\, \Omega}{d - 2}\, \pdiff{}{z} + 1 \right)\,
    \frac{ \tilde{K}_\mu{}^\rho - \delta_\mu{}^\rho\, \tilde{K} }{\Omega^{d-1}}
\nonumber \\
  &\hspace*{3.0truecm}
  + \frac{d - 1}{d - 2}\, \frac{ \tilde{g}_{\mu\nu} }{ \Omega^{d} }\,
    \left\{ L^2\, \Omega^2\, \left( \frac{\Omega'}{\Omega} \right)'
      + 1 - \frac{d - 2}{2}\, \left( 1 - L\, \Omega' \right)^2
    \right\}
  + \cdots
  \, \Bigg]
  ~.
\label{eq:exp-calT_lower-by_K}
\end{align}
The first term of the right-hand side of eq.~(\ref{eq:exp-calT_lower-by_K}) vanishes due to 
bulk equation of motion (\ref{eq:EOM-holo_semiC-Einstein_eqn-bulk}) if there are no matter fields. For the case of 
asymptotically AdS$_4$~($d = 3$) bulk spacetime, eq.~(\ref{eq:exp-calT_lower-by_K}) is expressed in a simple form. 
In fact, in this case the metric $\tilde{g}_{\mu\nu}$ has the following fall-off behavior toward the boundary $z \to 0$~\cite{Skenderis:2002wp}: 
\begin{align}
      & \tilde{g}_{\mu\nu}
  \sim \left\{ \bdyg_{\mu\nu} + \tilde{g}^{(2)}_{\mu\nu} \, \left( \frac{z}{L} \right)^2 + \cdots \right\}
  + \left\{ \tilde{g}^{(3)}_{\mu\nu}\, \left( \frac{z}{L} \right)^3 + \cdots \right\} 
  ~.
\label{eq:indcg-fall_off-expectation-d3}
\end{align}
The two curly brackets describe, respectively, two linearly independent solutions of 
eq.~(\ref{eq:EOM-holo_semiC-Einstein_eqn-bulk}) evaluated locally near $z=0$, so that  
$\tilde{g}^{(2)}_{\mu\nu}$ is determined by $\bdyg_{\mu\nu}$, while $\tilde{g}^{(3)}_{\mu\nu}$ 
is independent of $\bdyg_{\mu\nu}$ (as far as the boundary conditions (iii) are not imposed).  
In what follows, we call the first bracket {\em slow mode} and the second {\em fast mode}. 
Then, if $L\, \Omega(z)/z$ can be Taylor-expanded as $L\, \Omega/z = 1 + O(z^2)$, eq.~(\ref{eq:exp-calT_lower-by_K}) reduces to: 
%
\begin{align}
  & \pExp{\calT_{\mu\nu}}_{d=3}
  = \frac{3}{ 16\, \pi\, G_{4}\, L }~
    \left( \tilde{g}^{(3)}_{\mu\nu}
      - \bdyg_{\mu\nu}\, \tilde{g}^{(3)}_{\rho}{}^\rho \right)
  ~.
\label{eq:exp-calT_lower-by_K-d3-all}
\end{align}

\end{itemize}

\subsection{Perturbation analysis of the holographic semiclassical problem}
 
We study boundary dynamics by considering linear perturbations of the holographic semiclassical Einstein equations (\ref{eq:EOM-holo_semiC-Einstein_eqn-all}) for AdS gravity coupled to CFTs.

\subsubsection{Background}
Let us consider, as our background geometry, AdS$_{d+1}$ bulk with the curvature radius $L$, which satisfies the bulk Einstein equation (\ref{eq:EOM-holo_semiC-Einstein_eqn-bulk}), and AdS$_d$ boundary with the curvature radius $\ell$. 
In this background, it is convenient to take the following coordinate system, 
\begin{subequations}
\label{eq:bg_metric-all}
\begin{align}
  &\bar{ d s}_{d+1}^2
  = \Omega^{-2}(z)\, \left( dz^2 + \bar{g}_{\mu\nu}(x)\, dx^\mu dx^\nu \right)
& & \left( 0 < z/\ell < \pi \right)
  ~,
\label{eq:bg_metric} \\
  & \Omega(z) = \frac{\ell}{L}\, \sin\dfrac{z}{\ell}
  ~.
\label{eq:bg-Omega}
\end{align}
\end{subequations}
where $\bar{g}_{\mu\nu}(x)$ denotes the background AdS$_d$ metric with the curvature length $\ell$. 
In what follows, we express background quantities with \lq\lq\, $\Bar{~~}$\, \rq\rq\, 
so that, e.g., the covariant derivative compatible with the background metric $\bar{g}_{\mu \nu}$ is denoted by $\bar{D}_\mu$. 
Note that in the limit of either $z \to 0$, or $z/\ell \to \pi$, the metric (\ref{eq:bg_metric}) approaches AdS$_d$ boundary, 
and in the following, we take, without loss of generality, the limit $z = 0$ to consider our boundary semiclassical Einstein equations. 
As we did already above, we denote background quantities with {\em bar}, throughout our analysis. 

We note that since each $z$=const. hypersurface $\bar{\Sigma}_z$ with the background induced metric $\bar{g}_{\mu\nu}$ is totally geodesic, having 
the vanishing extrinsic curvature $\bar{K}_{\mu\nu} = 0$, eq.~(\ref{eq:exp-calT_lower-by_K}) reduces to 
\begin{align}
  & \pExp{\bar{\calT}_{\mu\nu}}
  = \lim_{z \to 0} \frac{ - (d - 1) }{ 4\, \pi\, G_{d+1}\, L }\,
  \left\{
  \bar{g}_{\mu\nu}\, \left( \frac{L}{2\, \ell} \right)^d\,
    \sin^{4-d}\frac{z}{2\, \ell}
  + \cdots \right\}
  ~,
\label{eq:bg_exp-calT_lower-by_K}
\end{align}
This implies that when $d$ is odd, without anomaly, conformal fields in pure AdS$_d$ boundary is in the conformal vacuum state, satisfying the semiclassical Einstein equations~(\ref{eq:EOM-holo_semiC-Einstein_eqn-bdy}). In particular, for $d = 3$, $\pExp{{\bar{\cal T}}_{\mu\nu}} = 0$. 
%

\subsubsection{Linear perturbations of the holographic semiclassical Einstein equations}
On this background~(\ref{eq:bg_metric-all}) with $\Omega(z)$ given by (\ref{eq:bg-Omega}), we consider metric perturbations $\delta G_{M N}$ under the radial gauge $\delta G_{zM}=0$, so that  the nontrivial part of 
the perturbations are expressed as $\delta G_{\mu \nu}= \Omega^{-2}(z) h_{\mu \nu}(x,z)$, or equivalently the whole metric becomes 
\begin{eqnarray}
  ds_{d+1}^2
  &=& \Omega^{-2}(z)\, \left[\, dz^2
  + \tilde{g}_{\mu\nu}(x,z) dx^\mu dx^\nu\, \right]
  \nonumber \\
  &=& \left(\frac{\ell}{L}\, \sin\dfrac{z}{\ell}\right)^{-2}\, \left[\, dz^2
  + \left\{ \bar{g}_{\mu\nu}(x) + h_{\mu\nu}(x, z)
    \right\}\, dx^\mu dx^\nu\, \right]
  ~. 
\label{eq:pert_metric}
\end{eqnarray}
%

In our background, each $z = \text{constant}$ hypersurface $\bar{\Sigma}$ is maximally symmetric with the AdS$_d$ metric $\bar{g}_{\mu \nu}$, and can 
be used to decompose the perturbation variables $h_{\mu \nu}$ into three types a la cosmological perturbations~\cite{Kodama:1984ziu}: the tensor-type $h^{(2)}_{T\, \mu\nu}$ satisfying $\bar{D}^\nu h^{(2)}_{T\, \mu\nu} = h^{(2)}_{T\, \mu}{}^\mu = 0$, the vector-type $h^{(1)}_{T\, \mu} = 0$ satisfying $\bar{D}^\mu h^{(1)}_{T\, \mu} = 0$, and the scalar-type $(h_L, h_T^{(0)})$, with respect to hypersurfaces $\bar{\Sigma}$, so that the perturbation 
variables are described as 
\begin{align}
  & h_{\mu\nu} 
  = h_L\, \bar{g}_{\mu\nu} + \bar{P}_{\mu\nu}\, h^{(0)}_T
  + 2 \bar{D}_{(\mu} h_{T\, \nu)}^{(1)} + h^{(2)}_{T\, \mu\nu}
  ~, 
\label{eq:tensor_dec-barH_mu_nu}
\end{align} 
where we have introduced the traceless projection operator:
\begin{align}
  & \bar{P}_{\mu\nu}
  := \bar{D}_{(\mu} \bar{D}_{\nu)}
  - \frac{1}{d} {\bar{g}_{\mu\nu}} \, \bar{D}^2
  ~.
\label{eq:def-barcalP}
\end{align}
Here and hereafter, the indices are raised and lowered by the background metric $\bar{g}^{\mu\nu}$ and $\bar{g}_{\mu\nu}$, and $h := h_\mu{}^\mu$ denotes the trace.  

Note that on a compact Riemannian Einstein manifold, any second-rank symmetric tensor field 
can uniquely be decomposed in this way~\cite{Ishibashi:2004wx}. However, such a tensor decomposition result, in particular, the uniqueness, would not be the case for our present setting with $(\bar{\Sigma}, \bar{g}_{\mu \nu})$ now being a Lorentzian submanifold. In this paper we proceed our analysis by assuming that the tensor decomposition results (\ref{eq:tensor_dec-barH_mu_nu}) hold. 
For convenience, the perturbation quantities appearing in our holographic semiclassical Einstein equations (\ref{eq:EOM-holo_semiC-Einstein_eqn-all}) are summarized in Appendix~\ref{App:C:pert} [see eqs.(\ref{A-eq:delta-E_MN-rad_gauge-all})--(\ref{A-eq:holo_semiC-Einstein_eq-pert-decomp-pre})].

In what follows, we consider each tensorial type of perturbations separately. Since we will 
apply the following perturbation formulas to the $d=3$ case in the subsequent sections, for convenience, 
we provide them by setting $d=3$ below. Generalizing the formulas to the general $d$ case should be straightforward~(Some of the key formulas are given in general dimension $d$ in Appendix~\ref{sec:formula-holo-semiC_Ein_eq}).   

\bigskip 


\noindent 
{\bf Tensor-type perturbation $h^{(2)}_{T \mu \nu}$} 

{}

\noindent 

\noindent 
The bulk Einstein equations~(\ref{eq:EOM-holo_semiC-Einstein_eqn-bulk}) and the mixed boundary condition 
(\ref{eq:EOM-holo_semiC-Einstein_eqn-bdy}) reduce, respectively, to 
\begin{subequations}
\label{eq:perturbe-EOM_grav-tensorP-sum-all}
\begin{align}
  {}
  & \partial^2_z h^{(2)}_{T\, \mu\nu}
  - 2\, \frac{\Omega'}{\Omega}\, \partial_z h^{(2)}_{T\, \mu\nu}
  + \left( \bar{D}^2 + \frac{2}{\ell^2} \right)\,
    h^{(2)}_{T\, \mu\nu} =0
  ~,
\label{eq:perturbe-EOM_grav-tensorP-sum_bulk}
\intertext{and}
  & - \frac{1}{\pi}\, \left( \frac{L}{\ell} \right)^3\,
    \left( \ell^2\, \bar{D}^2 + 2 \right)\,h_{T\, \mu\nu}^{(2)}
    \, \bigg|_{\text{$z \to 0$}}
  = \gamma_3 \times
    \frac{ L\, \partial_z h_{T\, \mu\nu}^{(2)} }{ \Omega^{2} }
    \, \bigg|_{\text{fast mode}}
  ~, 
\label{eq:perturbe-EOM_grav-tensorP-sum_bdy}
\end{align}
\end{subequations}
where the right-hand side of (\ref{eq:perturbe-EOM_grav-tensorP-sum_bdy}) should be evaluated 
in terms only of the fast mode of (\ref{eq:indcg-fall_off-expectation-d3}), and also where 
only terms relevant in $d<4$ are expressed. 
Here we have introduced the following dimensionless parameter: 
\begin{align}
  & \gamma_3
  := \frac{G_3}{G_{4} }\,  \frac{L^2}{\pi \ell} ~.
\label{eq:def-Upsilon}
\end{align}
Note that as we will see later, $\gamma_3$ express how large the effect of dual CFT energy-stress $\pExp{\calT_{\mu\nu}}$ is with respect to that of boundary cosmological constant $\Lambda_3/(8 \pi\, G_3)$.

\bigskip 

\noindent 
{\bf Vector-type perturbation $h^{(1)}_{T\, \mu}$} 

{}

\noindent 

\noindent 
The bulk Einstein equations~(\ref{eq:EOM-holo_semiC-Einstein_eqn-bulk}) and the mixed boundary condition (\ref{eq:EOM-holo_semiC-Einstein_eqn-bdy}) 
reduce, respectively, to 
\begin{subequations}
\label{eq:perturbe-EOM_grav-vectorP-sum-all}
\begin{align}
  & 0
  = \left( \bar{D}^2 - \frac{2}{\ell^2} \right)\,
    \partial_z h^{(1)}_{T\, \mu}
  ~,
\label{eq:perturbe-EOM_grav-vectorP-sum_bulk-1} \\
  & 0
  = \pdiff{}{z}\, \Bigg(
    \frac{ \partial_{z} \bar{D}_{(\mu} h_{T\, \nu)}^{(1)} }
         { \Omega^{2} } \Bigg)
  ~,
\label{eq:perturbe-EOM_grav-vectorP-sum_bulk-2}
\intertext{and}
  & 0
  = \frac{ L\, \partial_{z} \bar{D}_{(\mu} h_{T\, \nu)}^{(1)} }
         { \Omega^{2} }
  ~.
\label{eq:perturbe-EOM_grav-vectorP-sum_bdy}
\end{align}
\end{subequations}

\bigskip 

\noindent 
{\bf Scalar-type perturbation $(h_L, h_T^{(0)})$}  
{}

\noindent 

\noindent 
In terms of the gauge-invariant variable, which is analogous to the cosmological curvature perturbation~\cite{Kodama:1984ziu}
\begin{align}
  & \Phi 
  := h_L - \frac{1}{3}\, \bar{D}^2\, h^{(0)}_T
  ~,
\label{eq:gauge_inv-scalar_GW_L}
\end{align}
the bulk Einstein equations~(\ref{eq:EOM-holo_semiC-Einstein_eqn-bulk}) and the 
mixed boundary condition (\ref{eq:EOM-holo_semiC-Einstein_eqn-bdy}) can be expressed, respectively, as 
\begin{subequations}
\label{eq:perturbe-EOM_grav-scalarP-sum-all}
\begin{align}
  & 0
  = \pdiff{}{z}\, \left( \frac{ \partial_z h_L }{\Omega} \right)
  ~,
\label{eq:perturbe-EOM_grav-scalarP-sum_bulk-1} \\
  & 0
  = \bar{D}_\mu \partial_z
  \left( \Phi + \frac{1}{\ell^2}\, h^{(0)}_T \right)
  ~,
\label{eq:perturbe-EOM_grav-scalarP-sum_bulk-2} \\
  & 0
  = \left( \bar{D}^2 - \frac{3}{\ell^2} \right)\,
   \partial_z \left( \frac{ \Phi }{\Omega'} \right)
  ~,
\label{eq:perturbe-EOM_grav-scalarP-sum_bulk-3} \\
  & 0
  = \bar{P}_{\mu\nu} \left\{ \partial^2_z
  - 2 \, \frac{\Omega'}{\Omega}\, \partial_z
  - \frac{1}{\ell^2} \right\}\, \Phi
  ~,
\label{eq:perturbe-EOM_grav-scalarP-sum_bulk-4}
\intertext{and }
  & \left( \bar{D}^2 - \frac{3}{\ell^2} \right) \Phi
    \, \bigg|_{\text{slow mode}}
  = 0
  ~,
\label{eq:perturbe-EOM_grav-scalarP-sum_bdy-1} \\
  & \frac{1}{3 \pi}\, \left( \frac{L}{\ell} \right)^3\,
    {\bar P}_{\mu\nu}\, \Phi \, \bigg|_{\text{slow mode}}
  = \gamma_3 \times {\bar P}_{\mu\nu}\,
    \frac{ L\, \partial_z \Phi }{ \Omega^{2} }
    \, \bigg|_{\text{fast mode}}
  ~.
\label{eq:perturbe-EOM_grav-scalarP-sum_bdy-2}
\end{align}
\end{subequations}
Note that eq.~(\ref{eq:perturbe-EOM_grav-scalarP-sum_bulk-4}) can be rewritten as 
\begin{align}
  & 0
  = \pdiff{}{z}\, \frac{ \Omega'^2 }{ \Omega^{2} }\,
    \pdiff{}{z}\, \frac{ \bar{P}_{\mu\nu}\, \Phi }{\Omega'}
  ~.
\label{eq:perturbe-EOM_grav-scalarP-by-mfhL-4-I}
\end{align}
%


\subsubsection{The dimensionless parameter $\gamma_d$ that controls boundary dynamics}

As can be seen from eqs.~(\ref{eq:perturbe-EOM_grav-tensorP-sum_bdy}) and (\ref{eq:perturbe-EOM_grav-scalarP-sum_bdy-2}), perturbations of the bulk Einstein equations involve the dimensionless parameter $\gamma_3$ defined by (\ref{eq:def-Upsilon}). 
This parameter can be generalized to the case of general dimension $d$ as
\begin{align}
  & \gamma_d 
  := \frac{G_d\, L}{ \pi\, G_{d+1} }\, \left( \frac{L}{\ell} \right)^{d-2}
  ~, 
\label{eq:def-Upsilon:in:d}
\end{align}
and can be derived from the following holographic consideration.  
Suppose that the boundary conformal field theory has $N_{\text{dof}}$ \lq\lq degrees of freedom.\rq\rq 
Since the boundary curvature length scale is $\ell$, we can estimate $\pExp{\calT_{\mu\nu}} \sim N_{\text{dof}}/\ell^{d}$. 
Then, the semiclassical Einstein equations~(\ref{eq:EOM-holo_semiC-Einstein_eqn-bdy}) relate $\bdyR \sim 1/\ell^2$ and $G_d\, \pExp{\calT_{\mu \nu}}\sim G_d\, N_{\text{dof}}/\ell^{d}$, implying  
$1/\ell^2 \sim G_d\, N_{\text{dof}}/\ell^{d}$, 
and therefore should involve a dimensionless parameter $\gamma_d \sim G_d\, N_{\text{dof}}/\ell^{d-2}$. From the AdS/CFT correspondence, we can also estimate that $N_{\text{dof}} \sim L^{d-1}/G_{d+1}$ 
and hence obtain the dimensionless parameter $\gamma_d \sim G_d\, N_{\text{dof}}/\ell^{d-2}$ from the relation  
\begin{align}
  & \frac{ G_d }{ \ell^{d-2} }\, \frac{ L^{d-1} }{ G_{d+1} }
  = \frac{ G_d\, L }{G_{d+1}}\, \left( \frac{L}{\ell} \right)^{d-2}
  = \pi\, \gamma_d
 ~.
\end{align}
  %
  %

Note that if the $d$-dimensional boundary contains, e.g., a black hole with the mass $M$, then we can introduce another dimensionless parameter $G_d\, M/\ell^{d-3} \sim (r_+/\ell)^{d-3}$, or $\gamma_{\rm BH} := \ell\, T_{\tBH} \sim r_+/\ell$. In such a case, the dimensionless parameter $\gamma_{\rm BH}$ specifies the Hawking-Page phase transition on the boundary. 
The parameter $\gamma_d \sim G_d\, N_{\text{dof}}/\ell^{d-2}$ is involved even when a holographic system under consideration has zero-temperature. 


One can also see that the parameter $\gamma_d$ represents the ratio of the strength of the boundary quantum stress-energy $\langle {\cal T}_{\mu \nu} \rangle$ with respect to that of the boundary cosmological constant $\Lambda_d/(8 \pi\, G_d)=-4\pi G_d(d-2)(d-1)/(8 \pi\, G_d \ell^2)$. 
From eq.~(\ref{A-eq:exp-calT_lower-by_K-pert-2-pre}), one can estimate $\delta\pExp{\calT_{\mu\nu}}$ roughly as
\begin{align}
 & \delta\pExp{\calT}
 \sim \frac{1}{ G_{d+1}\, L }\,
   \frac{ L\, \partial_z h }{ \Omega^{d-1} }
   \, \bigg|_{\text{fast mode}}
 \sim \frac{ h^{(f)} }{ G_{d+1}\, L }\,
   \left( \frac{L}{\ell} \right)^d
 ~, 
\label{eq:delta-exp-calT-order}
\end{align}
where we denote $h_{\mu\nu}(x, z)
  \sim \pbdyg_{\mu\nu}(x)\, \left( 1 + \cdots \right)
  + h^{(f)}_{\mu\nu}(x)\, (z/L)^d\, \left( 1 + \cdots \right)$. 
Then, on one hand, the response $\calK$ of the dual quantum field $\pExp{\calT_{\mu\nu}}$ against the perturbations of 
the boundary metric $\pbdyg_{\rho\sigma}$ can be estimated roughly as  
\begin{align}
    & \calK
    \sim \frac{ \delta \pExp{\calT} }{\pbdyg}
    \sim \frac{1}{ G_{d+1}\, L }\, \left( \frac{L}{\ell} \right)^d\,
      \frac{h^{(f)}}{\pbdyg}
    ~. 
  \label{eq:calK-order}
\end{align}
On the other hand, the contribution to the CFT stress-energy tensor from the cosmological constant 
$\Lambda_d = - (d - 1)\, (d - 2)/(2\, \ell^2)$ is estimated as 
  $\calT^\Lambda_{\mu\nu} = - \Lambda_d\, \bdyg_{\mu\nu}/(8 \pi G_d)$, 
and thus its response $\calK^\Lambda$ is
  \begin{align}
    & \calK^\Lambda
    \sim \frac{ \delta \pExp{\calT^\Lambda} }{\pbdyg}
    \sim \frac{\Lambda_d}{G_d}
    \sim \frac{1}{G_d\, \ell^2}
    ~.
  \label{eq:calK_Lambda-order}
  \end{align}
Therefore, assuming $h^{(f)}/\pbdyg = O(1)$, one can find the strength of $\pExp{\calT_{\mu\nu}}$ compared with $\Lambda_d$ becomes  
\begin{align}
    & \frac{\calK}{\calK^\Lambda}
    \sim \frac{G_d\, \ell^2}{ G_{d+1}\, L }\, \left( \frac{L}{\ell} \right)^d\,
      \frac{h^{(f)}}{\delta \cal{G}}
    \sim \frac{G_d\, L}{ G_{d+1} }\, \left( \frac{L}{\ell} \right)^{d-2}
    = \pi\, \gamma_d
    ~.
\label{eq:calK-calK_Lambda-order-ratio}
\end{align}

%


\newcommand{\tm}{\textrm}
\newcommand{\vp}{\varphi}
\newcommand{\blue}{\textcolor{blue}}
\newcommand{\yellow}{\textcolor{yellow}}
\newcommand{\red}{\textcolor{red}}
\newcommand{\green}{\textcolor{green}}
\newcommand{\magenta}{\textcolor{magenta}}
\newcommand{\cyan}{\textcolor{cyan}}
\newcommand{\ovl}[1]{\overline{#1}}
\newcommand{\wt}[1]{\widetilde{#1}}
\newcommand{\eq}[1]{Eq.~(\ref{#1})}
\newcommand{\eqn}[1]{(\ref{#1})}
\newcommand{\p}{\partial}
\newcommand{\pslash}{p\kern-1ex /}
\newcommand{\qslash}{q\kern-1ex /}
\newcommand{\lslash}{l\kern-1ex /}
\newcommand{\sslash}{s\kern-1ex /}
\newcommand{\kaslash}{k_a\kern-2ex /}
\newcommand{\kbslash}{k_b\kern-2ex /}
\newcommand{\Dslash}{{\cal D}\kern-1.5ex /}
\newcommand{\bpsi}{\overline{\psi}}
\newcommand{\bc}{\overline{c}}
\newcommand{\tr}{{\rm tr}}
\newcommand{\vev}[1]{\langle #1 \rangle}
\newcommand{\VEV}[1]{\left\langle{\rm T} #1\right\rangle}
\newcommand{\beqa}{\begin{eqnarray}}
\newcommand{\eeqa}{\end{eqnarray}}
\newcommand{\bq}{\mbox{\boldmath $q$}}
\newcommand{\bp}{\mbox{\boldmath $p$}}
\newcommand{\bk}{\mbox{\boldmath $k$}}
\newcommand{\br}{\mbox{\boldmath $r$}}
\newcommand{\bs}{\mbox{\boldmath $s$}}
\newcommand{\bt}{\mbox{\boldmath $t$}}
\newcommand{\bx}{\mbox{\boldmath $x$}}
\newcommand{\by}{\mbox{\boldmath $y$}}
\newcommand{\bz}{\mbox{\boldmath $z$}}
\newcommand{\bv}{\mbox{\boldmath $v$}}
\newcommand{\bw}{\mbox{\boldmath $w$}}
\newcommand{\bL}{\mbox{\boldmath $L$}}
\newcommand{\bJ}{\mbox{\boldmath $J$}}
\newcommand{\bP}{\mbox{\boldmath $P$}}
\newcommand{\bQ}{\mbox{\boldmath $Q$}}
\newcommand{\bS}{\mbox{\boldmath $S$}}
\newcommand{\bR}{\mbox{\boldmath $R$}}
\newcommand{\bsigma}{\mbox{\boldmath $\sigma$}}
\newcommand{\btau}{\mbox{\boldmath $\tau$}}
\newcommand{\brho}{\mbox{\boldmath $\rho$}}
\newcommand{\bxi}{\mbox{\boldmath $\xi$}}
\newcommand{\bns}{\mbox{\scriptsize \boldmath $n$}}
\newcommand{\brs}{\mbox{\scriptsize \boldmath $r$}}
\newcommand{\bss}{\mbox{\scriptsize \boldmath $s$}}
\newcommand{\bts}{\mbox{\scriptsize \boldmath $t$}}
\newcommand{\bxs}{\mbox{\scriptsize \boldmath $x$}}
\newcommand{\bys}{\mbox{\scriptsize \boldmath $y$}}
\newcommand{\bps}{\mbox{\scriptsize \boldmath $p$}}
\newcommand{\bqs}{\mbox{\scriptsize \boldmath $q$}}
\newcommand{\bks}{\mbox{\scriptsize \boldmath $k$}}
\newcommand{\bLs}{\mbox{\scriptsize\boldmath $L$}}
\newcommand{\bJs}{\mbox{\scriptsize\boldmath $J$}}
\newcommand{\bPs}{\mbox{\scriptsize\boldmath $P$}}
\newcommand{\bQs}{\mbox{\scriptsize\boldmath $Q$}}
\newcommand{\bSs}{\mbox{\scriptsize\boldmath $S$}}
\newcommand{\bRs}{\mbox{\scriptsize\boldmath $R$}}
\newcommand{\bsigmas}{\mbox{\scriptsize \boldmath $\sigma$}}
\newcommand{\btaus}{\mbox{\scriptsize \boldmath $\tau$}}
\newcommand{\brhos}{\mbox{\scriptsize \boldmath $\rho$}}
\newcommand{\bxis}{\mbox{\scriptsize \boldmath $\xi$}}
\renewcommand{\Re}{\mathrm{Re}\,}
\renewcommand{\Im}{\mathrm{Im}\,}
\newcommand{\bpm}{\begin{pmatrix}}
\newcommand{\epm}{\end{pmatrix}}
\newcommand{\bbm}{\begin{bmatrix}}
\newcommand{\ebm}{\end{bmatrix}}
\newcommand{\sla}[1]{{\ooalign{\hfil/\hfil\crcr$#1$}}}
\newcommand{\plaq}[1]{U_{{\rm P},#1}}  

\def\n{\nonumber}
\def\p{\partial}
\def\u{\underline}
\newcommand{\Exp}[1]{\left\langle~#1~\right\rangle}

\def\Tr{\hbox{Tr}}
\newcommand{\be}{\begin{equation}}
\newcommand{\ee}{\end{equation}}
\newcommand{\bea}{\begin{eqnarray}}
\newcommand{\eea}{\end{eqnarray}}
\newcommand{\beas}{\begin{eqnarray*}}
\newcommand{\eeas}{\end{eqnarray*}}
\newcommand{\nn}{\nonumber}
\font\cmsss=cmss8
\def\C{{\hbox{\cmsss C}}}
\font\cmss=cmss10
\def\bigC{{\hbox{\cmss C}}}
\def\scriptlap{{\kern1pt\vbox{\hrule height 0.8pt\hbox{\vrule width 0.8pt
  \hskip2pt\vbox{\vskip 4pt}\hskip 2pt\vrule width 0.4pt}\hrule height 0.4pt}
  \kern1pt}}
\def\ba{{\bar{a}}}
\def\bb{{\bar{b}}}
\def\bc{{\bar{c}}}
\def\bphi{{\Phi}}
\def\Bigggl{\mathopen\Biggg}
\def\Bigggr{\mathclose\Biggg}
\def\Biggg#1{{\hbox{$\left#1\vbox to 25pt{}\right.\n@space$}}}
\def\n@space{\nulldelimiterspace=0pt \m@th}
\def\m@th{\mathsurround = 0pt}

\section{Semiclassical dynamics in $3$-dimensional black holes}
\label{sec:3} 
In this section, we study perturbative dynamics of $3$-dimensional boundary gravity as an application of our holographic 
semiclassical Einstein equations (\ref{eq:EOM-holo_semiC-Einstein_eqn-all}). 
As a solution of the $4$-dimensional vacuum bulk Einstein equations (\ref{eq:EOM-holo_semiC-Einstein_eqn-bulk}), 
\begin{align}
\label{bulk_equation}
R_{MN}=-\frac{3}{L^2}\,G_{MN},    
\end{align}
we start with the AdS$_4$ bulk with AdS$_3$ boundary as our background geometry,  
\begin{align} 
\label{metric_BTZ_BS}
& ds_4^2=\Omega^{-2}(z) \left\{ dz^2 + \bar{g}_{\mu\nu}(x)dx^\mu dx^\nu \right\} \,, \quad 
\Omega=\frac{\ell}{L}\sin\left(\frac{z}{\ell} \right) \,,  
\nonumber \\ 
& \bar{ds}_3^2=\bar{g}_{\mu\nu}(x)dx^\mu dx^\nu
=-\left(\frac{r^2}{\ell^2}-M\right)dt^2+\dfrac{dr^2}{ \dfrac{r^2}{\ell^2}-M}+r^2d \varphi^2 \,.   
\end{align}
Note that in the above expression of the AdS$_3$ metric, we adopt the coordinates $x^\mu=(t,r,\varphi)$ 
so that $\bar{ds}^2_3= \bar{g}_{\mu\nu}dx^\mu dx^\nu$ can manifestly be seen as the covering space of the static BTZ black hole~\cite{Banados:1992wn} 
with the length scale $\ell$. Our boundary metric $\bar{g}_{\mu\nu}$ satisfies 
(\ref{eq:EOM-holo_semiC-Einstein_eqn-bdy}) with 
$\langle {\cal T}_{\mu \nu } \rangle=0$, namely, 
\begin{equation}
 {\cal R}_{\mu \nu} = -\dfrac{2}{\ell^2}{\bar g}_{\mu \nu} \,,  
\end{equation} 
and accordingly we refer to our background boundary AdS$_3$ as the {\em BTZ black hole with vanishing expectation value of the stress-energy tensor}. In this respect, our bulk metric $ds_4^2$ can be viewed as the metric for (the covering space of) AdS black string.


\subsection{Perturbed semiclassical Einstein equations}
\label{sec:pert:semiEineqn}

Now let us consider perturbation of the solution~(\ref{metric_BTZ_BS}). We restrict our attention to the tensor-type perturbation, $h^{(2)}_{T \mu \nu}$, which satisfies the transverse-traceless condition 
\begin{align}
\label{transverse-traceless}
\bar{D}^\mu h^{(2)}_{T \mu\nu}=\bar{g}^{\mu\nu} h^{(2)}_{T \mu\nu}=0 \,. 
\end{align}
Then, from the Einstein equations~(\ref{bulk_equation}), we obtain (\ref{eq:perturbe-EOM_grav-tensorP-sum_bulk}), i.e., 
\begin{align} 
\label{pert_h_munu}
h^{(2)}{}''_{T \mu\nu}+\bar{D}^2h^{(2)}_{T \mu\nu}-2\dfrac{\Omega'}{\Omega} h^{(2)}{}'_{T\mu\nu}+\frac{2}{\ell^2}h^{(2)}_{T \mu\nu}=0 \,, 
\end{align}
where here and in the following the {\em prime} denotes the derivative with respect to $z$. 
By making the following ansatz for separation of variables 
\begin{equation}
h^{(2)}_{T \mu\nu}(x,z)=\Hat{\xi}(z)H_{\mu\nu}(x) \,,
\label{ans:h}
\end{equation} 
we decompose eq.~(\ref{pert_h_munu}) into 
the $3$-dimensional part 
\begin{align} 
\label{Basic_eq_H_munu}
\bar{D}^2H_{\mu\nu}+\frac{2}{\ell^2}H_{\mu\nu}=m^2H_{\mu\nu}, 
\end{align}
and the radial part
\begin{align}
\label{Basic_eq_xi}
  0 = \left( \odiffII{}{z} - 2\, \frac{ \Omega'(z) }{\Omega}\, \odiff{}{z}
  + m^2 \right)\, \Hat{\xi}
  = \left( \odiffII{}{z} - \frac{2}{\ell\, \tan(z/\ell)}\, \odiff{}{z} + m^2
  \right)\, \Hat{\xi}~,
\end{align}
with a separation constant $m^2$. 

For convenience, we introduce the following coordinate $w$,  
\begin{align} 
\label{FG_z_w_relation}
z=w\left(1-\frac{1}{12 \ell^2}w^2+\frac{1}{80 \ell^4}w^4+\cdots\right), 
\end{align} 
so that the metric reduces to the Fefferman-Graham form
\begin{align} 
\label{FG_coordinate}
& ds_4^2=\frac{L^2}{w^2}(dw^2+ \tilde{g}_{\mu\nu}(x,w)dx^\mu dx^\nu), \nonumber \\
& \tilde{g}_{\mu\nu}(x,w)=g_{(0)\mu\nu}(x)+w^2g_{(2)\mu\nu}(x)+w^3g_{(3)\mu\nu}(x)+\cdots \,. 
\end{align}
Note that even if we expand ${\tilde g}_{\mu \nu}$ in terms of $z$ instead of $w$, we still obtain the same coefficients $g_{(0) \mu \nu}(x),..., $ until at $O(z^3)$ due to eq.~(\ref{FG_z_w_relation}).  

As shown in~\cite{deHaro:2000vlm}, the coefficient $g_{(2)\mu \nu}$ is given by   
\begin{align} 
\label{coe_g2}
g_{(2)\mu\nu}=-{\cal R}_{\mu\nu}+\frac{1}{4}{\cal R}g_{(0)\mu\nu} \,, 
\end{align}
where ${\cal R}_{\mu\nu}$ and ${\cal R}$ are the Ricci tensor and Ricci scalar for the boundary metric $g_{(0)\mu\nu}$. Since the perturbation of ${\cal R}_{\mu \nu}$ is written as  
\begin{equation}
\label{delta:calRicci}
 \delta {\cal R}_{\mu \nu} 
  = -\dfrac{1}{2}\left( \bar{D}^2 + \dfrac{6}{\ell^2}\right) \delta g_{(0)\mu \nu} 
  = - \left(\dfrac{m^2}{2} + \dfrac{2}{\ell^2} \right) \delta g_{(0)\mu \nu} \,, 
\end{equation}
under the condition~(\ref{Basic_eq_H_munu}), we obtain  
\begin{align}
\label{pert_ricci}
& \delta {\cal R}=\bar{g}_{(0)}^{\mu\nu}\delta {\cal R}_{\mu\nu} = 0 \,, 
\nonumber \\
& \delta {\cal R}_{\mu\nu}=-\delta g_{(2)\mu\nu}-\frac{3}{2 \ell^2}\delta g_{(0)\mu\nu} 
 \,,  
\end{align}
where we have used the fact that 
\begin{align}
\delta {g_{(n)}}^\mu{}_\mu = \delta {g_{(n)}}_{\mu\nu} \bar{g}^{(0)\mu\nu} = 0 \,, 
\end{align}
at any order $n$ for the linearized perturbations under our metric ansatz (\ref{ans:h}), and eqs.~(\ref{coe_g2}) and (\ref{delta:calRicci}). 
Then, we can reduce the linearized 
semiclassical Einstein equations~(\ref{eq:EOM-holo_semiC-Einstein_eqn-bdy}) to   
\begin{align}
\label{linearized_semi}
-{\delta g}_{(2)\mu\nu}+\frac{{\delta g}_{(0)\mu\nu}}{2 \ell^2}
=8\pi G_3\delta\Exp{{\cal T}_{\mu\nu}} = \dfrac{3}{2}\pi \ell \gamma_3 {\delta g}_{(3)\mu\nu} \,,  
\end{align}
where we have used the formula~\cite{deHaro:2000vlm}
\begin{align}
\label{stress_energy_tensor_formula} 
\Exp{\cal{T}_{\mu\nu}} = \frac{3L^2}{16\pi G_4}g_{(3)\mu\nu} \,.  
\end{align}

Now note that under our ansatz~(\ref{ans:h}), both $\delta{g}_{(0)\mu\nu}$ and $\delta{g}_{(3)\mu\nu}$ are proportional to $H_{\mu \nu}(x)$. Then, by expanding the solution of (\ref{Basic_eq_xi}) as 
\begin{eqnarray}
\label{expand_xi}
  \Hat{\xi} = a_0+a_1z+a_2z^2 + a_3 z^3+\cdots \,, 
%
\end{eqnarray}
and using eqs.~(\ref{FG_z_w_relation}) and (\ref{FG_coordinate}), we can reduce eq.~(\ref{linearized_semi}) to the following simple relation,  
%
\begin{align} 
\label{relation_a0_a3}
a_3=-\frac{{m}^2}{3\pi \ell \gamma_3}a_0 \,. 
\end{align}
%

Under the coordinate transformation, $\eta:=(1-\cos(z/\ell))/2$, eq.~(\ref{Basic_eq_xi}) becomes 
\begin{align}
\label{eta_eq} 
  \eta\, (1 - \eta)\, \odiffII{ \Hat{\xi} }{\eta}
  + \left( - \frac{1}{2} + \eta \right)\, \odiff{ \Hat{\xi} }{\eta}
  + \hat{m}^2\, \Hat{\xi} = 0~,
%
\end{align}
where here and hereafter 
\begin{equation}
\hat{m}^2:=m^2\ell^2 \,. 
\end{equation} 
The solution is given in terms of the hypergeometric function as 
\begin{subequations}
\begin{align}
  & \Hat{\xi}
  = c_1\, F\Big( -1 + p,\, -1 - p,\, - \frac{1}{2}\,;\, 1 - \eta \Big)
  + c_2\, (1 - \eta)^{3/2}\,
    F\Big( \frac{1}{2} + p,\, \frac{1}{2} - p,\, \frac{5}{2}\,;\, 1 - \eta
     \Big)
  ~,
\label{eq:sol-hxi} \\
  & p := \sqrt{1+\hat{m}^2}~.
\label{eq:def-p}
\end{align}
\end{subequations}
We need to impose boundary conditions at the rest of the AdS boundary $\partial M \setminus \Sigma_0$, depicted in Figure~\ref{geom}, which in the present coordinates correspond to $\eta=1$~($z=\pi \ell$). We choose the Dirichlet condition $\Hat{\xi}=0$ at $\eta=1$~($z=\pi \ell$), and thus 
$c_1$ should be zero. 
By using Gaussian transform formula, 
\begin{align} 
\label{Gaussian transform formula}
& F(a,b,c;1-\eta)
=\frac{\Gamma(c)\Gamma(a+b-c)}{\Gamma(a)\Gamma(b)}\eta^{c-a-b}
F(c-a,c-b,c-a-b+1;\eta) \nonumber \\
& \qquad \qquad \qquad \quad {} 
  +\frac{\Gamma(c)\Gamma(c-a-b)}{\Gamma(c-a)\Gamma(c-b)}
F(a,b, a+b-c+1;\eta) \,, 
\end{align}
the asymptotic behavior of $\xi$ at the AdS boundary $z=0$ becomes  
\begin{align} 
\label{xi_asymptotic_sol}
 \frac{\Hat{\xi}}{c_2}
  &= (1 - \eta)^{3/2}\, \times
  \bigg\{
  \frac{ \Gamma(5/2)\, \Gamma(- 3/2) }{ \Gamma(1/2 - p)\, \Gamma(1/2 + p) }\,
  \eta^{3/2}\, F\Big( 2 + p,\, 2 - p,\, \frac{5}{2}\,;\, \eta \Big)
\nonumber \\
  &\hspace*{3.5truecm}
  + \frac{ \Gamma(5/2)\, \Gamma(3/2) }{ \Gamma(2 + p)\, \Gamma(2 - p) }\,
    F\Big( \frac{1}{2} - p,\, \frac{1}{2} + p,\, - \frac{1}{2}\,;\,\eta \Big)
  \bigg\}
\nonumber \\
  &\sim \frac{ - 3\, \sin(\pi\, p) }{ 8\, p\, (p^2 - 1) }\,
  \left\{ 1 + \frac{p^2 - 1}{2}\, \frac{z^2}{\ell^2}
  - \frac{ p\, (p^2 - 1) }{ 3\, \tan(\pi\, p) }\, \frac{z^3}{\ell^3} + \cdots
  \right\}~.
%
\end{align}
From eq.~(\ref{relation_a0_a3}), one obtains 
\be 
\label{bulk_eigenvalue}
\gamma_3=\frac{\tan\pi\sqrt{1+\hat{m}^2}}{\pi\sqrt{1+\hat{m}^2}} \,. 
\ee
When $\hat{m}^2>0$, there is an infinite number of solutions, $\hat{m}$ satisfying eq.~(\ref{bulk_eigenvalue}) for any $\gamma_3>0$. On the other hand, when $\hat{m}^2<0$, the solution $\hat{m}$ should be bounded 
from below by $-1<\hat{m}^2$, otherwise, the BF bound~\cite{Breitenlohner:1982jf,Breitenlohner:1982bm} on the boundary is violated, as shown in the next 
section. In the range of $-3/4\le \hat{m}^2\le 0$, there is no solution for any positive $\gamma_3$. 
As a result, there is only one solution, $\hat{m}$ in the range $-1<\hat{m}^2<-3/4$ for any $\gamma_3>1$.


\subsection{Perturbed solutions}

In this subsection, we first construct analytic solutions of eq.~(\ref{Basic_eq_H_munu}) with (\ref{bulk_eigenvalue}) 
by considering static perturbations. We also discuss physical implication of the dimensionless parameter 
$\gamma_3$ given by eq.~(\ref{bulk_eigenvalue}) above. 
Then, we derive formulas for time-dependent perturbations, which will be used subsequently to discuss boundary dynamics.

\subsubsection{Static perturbations} 
Having our background metric~(\ref{metric_BTZ_BS}), we find it more convenient to use the following Eddington-Finkelstein form for the boundary metric:   
\begin{align} 
\label{v_r_coor}
& ds_3^2=-f(1-\epsilon {\cal H}_{vv}(r))dv^2+2(1+\epsilon {\cal H}_{vr}(r))dvdr \nonumber \\
&+\epsilon {\cal H}_{rr}(r)dr^2+r^2(1+\epsilon {\cal H}_{\varphi \varphi}(r))d\varphi^2, \quad f=\frac{r^2}{\ell^2}-M \,,
\end{align}
where $\epsilon$ is an infinitesimally small parameter. 
From 
eqs.~(\ref{Basic_eq_H_munu}) and (\ref{v_r_coor}), we find that ${\cal H}_{vv}$ can be regarded as our master variable, by which ${\cal H}_{\varphi  \varphi}, {\cal H}_{vr}$, and ${\cal H}_{rr}$ are expressed as 
\begin{align} 
\label{static_HvrKHrr}
& {\cal H}_{\varphi \varphi} =-2{\cal H}_{vr} - f{\cal H}_{rr} \,, \quad {\cal H}_{vr}=-{\cal H}_{vv} \,, 
\quad {\cal H}_{rr} = \dfrac{\ell^2}{r_h^2}\dfrac{(1+\hat{m}^2)u{\cal H}_{vv} + 2 u^3 {\cal H}_{vv}' }{ 1+\hat{m}^2(1-u)} \,, 
%
\end{align}
and that ${\cal H}_{vv}$ obeys 
\begin{align} 
\label{master_static}
& (1-u)[1+\hat{m}^2(1-u)] {\cal H}_{vv}''- [ 2+ \hat{m}^2(1-u)] {\cal H}_{vv}' 
  - \dfrac{\hat{m}^2}{4u^2}[1+2 u + \hat{m}^2(1- u)] {\cal H}_{vv}=0 \,,
%
\end{align}
where we have introduced the new variable $u := r_h^2/r^2$ with $r_h:= \ell \sqrt{M}$ being the background horizon radius, 
and here and hereafter the {\em prime} denotes the derivative with respect to $u$.   

The asymptotic behavior of ${\cal H}_{vv}$ in eq.~(\ref{master_static}) is given by  
\begin{align}
\label{asymptotic_static} 
{\cal H}_{vv}\simeq c_1 u^{(1-p)/2} + c_2 u^{(1+p)/2} \,.  
\end{align} 
This asymptotic behavior is the same as that of a massive scalar field $\phi$ with mass-squared $m^2$ 
on the background of AdS$_3$. 
We are not interested in any unstable solution with $\hat{m}^2<-1$, which violates the BF bound~\cite{Breitenlohner:1982jf,Breitenlohner:1982bm}, and therefore we will focus on the case 
$-1 \leq \hat{m}^2$ (and hence $p \geq 0$). When $\hat{m}^2=0$ ($p=1$), a logarithmic mode appears, but it is not permitted by eq.~(\ref{bulk_eigenvalue}) unless $\gamma_3 = 0$~($G_3=0$). 
When $-1<\hat{m}^2<0$ ($0<p<1$), there are two normalizable modes. Such a scalar field model is effectively derived from $3$-dimensional massless conformal scalar field in the background of asymptotically AdS$_3$ spacetime.  
The effective negative mass-squared becomes $\hat{m}^2=-3/4$, 
and there is an exact static black hole solution which includes two 
normalizable modes~\cite{Martinez:1996gn}. So, we do not impose any particular boundary condition 
in the mass range, $-1<\hat{m}^2<0$. When $\hat{m}^2>0$ ($1<p$), we choose the Dirichlet boundary 
condition, $c_1=0$ so that the perturbed field does not diverge at the AdS boundary. 
We also impose the regularity condition at the horizon, $u=1$. By eq.~(\ref{master_static}), it is given by  
\begin{align} 
\label{seisoku_con}
 {\cal H}_{vv}' (1)=-\frac{3\hat{m}^2}{8}{\cal H}_{vv}(1) \,.
\end{align}

Now, under our holographic setting given in the previous subsection~\ref{sec:pert:semiEineqn}, we obtain the following theorem. 

\medskip 

\noindent 
{\it Theorem}: There is no static, regular asymptotically AdS$_3$ boundary black hole as a solution to the holographic perturbed semiclassical Einstein equations with non-vanishing stress-energy tensor, when $\gamma_3 <1$. 

\medskip 

\noindent 
{\it Proof}: First, let us consider the case $\hat{m}^2>0$ ($p>1$). Since eq.~(\ref{master_static}) is a linear equation, we can take 
${\cal H}_{vv}(1)=1$ without loss of generality. From the asymptotic boundary condition, $c_1=0$, ${\cal H}_{vv}$ must converge 
to zero at $u=0$ by eq.~(\ref{asymptotic_static}). As ${\cal H}_{vv}'(1)<0$ by eq.~(\ref{seisoku_con}), ${\cal H}_{vv}$ 
would have a local maximum at some $u=u_{\rm m}~(0<u_{\rm m}<1)$ between the horizon and the AdS boundary. This means 
that ${\cal H}_{vv}'(u_{\rm m})=0$ and ${\cal H}_{vv}''(u_{\rm m})<0$. This is impossible because 
\begin{align}
(1-u_{\rm m})(1+\hat{m}^2-\hat{m}^2 u_{\rm m})>0 \,, \quad 
 \dfrac{\hat{m}^2}{4u_{\rm m}^2}[1+2 u_{\rm m} + \hat{m}^2(1- u_{\rm m})]>0 \,. 
\end{align}   
Next, let us consider the case $\hat{m}^2<0$. Then, it should be $-1<\hat{m}^2<-3/4$ by eqs.~(\ref{bulk_eigenvalue}) and 
(\ref{asymptotic_static}). As shown in the previous section,  
the right-hand side of eq.~(\ref{bulk_eigenvalue}) is bounded by $1$ from below. 
This is impossible when $\gamma_3<1$. 
 $\Box$

\medskip 
This theorem states that there is a lower bound 
for the three dimensional gravitational constant $G_3$, in order for a semiclassical static black hole solution with non-zero expectation value of the stress-energy tensor to exist. By eq.~(\ref{eq:def-Upsilon}) 
or equivalently (\ref{eq:def-Upsilon:in:d}), the lower bound is given by 
\begin{align}
\label{lower_bound_G3}
G_{3, {\rm min}}=\frac{\pi \ell}{L^2}G_4 \,, 
\end{align} 
which is equivalent to $\gamma_3 =1$. 
As shown below, the existence of the static black hole solution suggests that the BTZ black hole with 
$\Exp{{\cal T}_{\mu\nu}}=0$ is unstable when $G_3>G_{3, {\rm min}}$ ($\gamma_3>1$). On the other hand, when $G_3<G_{3, {\rm min}} (\gamma_3<1)$, 
the backreaction of non-zero $\Exp{{\cal T}_{\mu\nu}}$ through the semiclassical eq.~(\ref{linearized_semi}) is also small, 
and the BTZ black hole with $\Exp{{\cal T}_{\mu\nu}}=0$ would be stable. 
From the bulk point of view, the boundary condition at the AdS boundary is a mixed boundary condition satisfying eq.~(\ref{relation_a0_a3}). So, when $G_3$ is small enough, $|a_0|$ approaches zero for a 
fixed $a_3$, and hence, the boundary condition approaches usual Dirichlet boundary condition. This agrees with 
the result that the BTZ black hole is stable against vacuum perturbations under the Dirichlet 
boundary condition~\cite{Liu:2008ds, Marolf:2019wkz}. It is noteworthy that $G_3$ becomes a control parameter which determines the stability of the BTZ solution. 
 
\begin{figure}[htbp]
  \begin{center}
  \includegraphics[width=100mm]{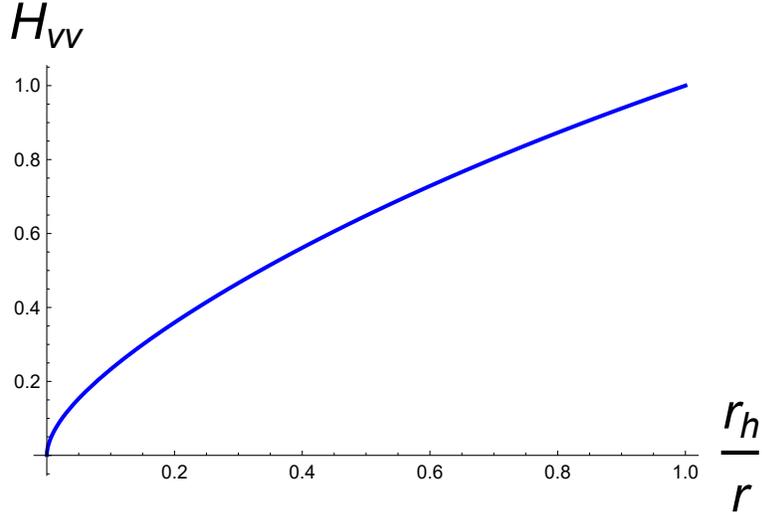}
  \end{center}
  \caption{${\cal H}_{vv}$, normalized at the horizon $u=1(r=r_h)$, is shown for $\hat{m}^2=-4/5$.}
  \label{static_semiclassical_BH}
 \end{figure} 
\begin{figure}[htbp]
  \begin{center}
  \includegraphics[width=100mm]{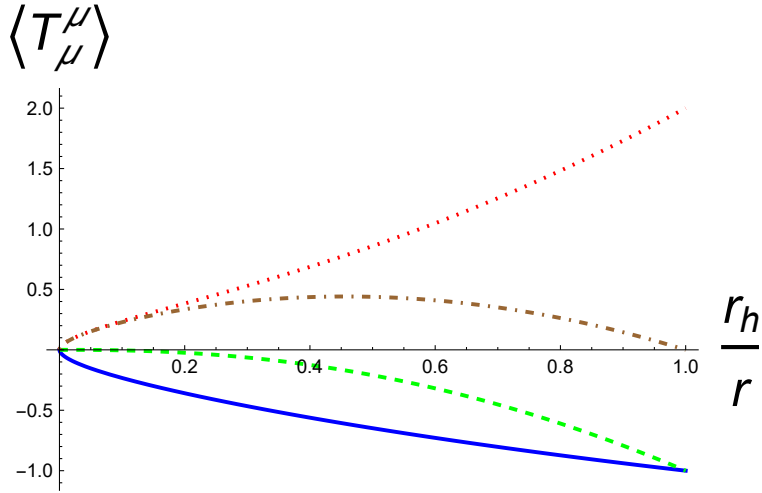}
  \end{center}
  \caption{$\langle{\cal T}_t{}^t \rangle $~(solid, blue), $\langle{\cal T}_r{}^r \rangle $~(dashed, green), $\langle{\cal T}_{\varphi}{}^\varphi \rangle$~(dotted, red), and 
  $f{\cal H}_{rr}$~(dot-dashed, brown) are shown under the normalization by the absolute value of 
  $\langle {\cal T}_t{}^t \rangle$ at the horizon for $\hat{m}^2=-4/5$.}
  \label{stress_energy_fig}
 \end{figure} 
 
When $-1<\hat{m}^2<-3/4$, one obtains the analytic solution from eqs.~(\ref{static_HvrKHrr}) and (\ref{master_static}), under the boundary condition (\ref{seisoku_con}). By introducing a new variable 
\begin{equation}
\label{def:Psi}
\Psi:={\cal H}_{vv}-f{\cal H}_{rr} \,,
\end{equation}
${\cal H}_{vv}$, ${\cal H}_{rr}$, 
and ${\cal H}_{\varphi \varphi}$ are expressed via eqs.~(\ref{static_HvrKHrr}) and (\ref{master_static}) as 
\begin{align}
\label{static_Phi}
& {\cal H}_{vv}= \left(\frac{3}{u}-2\right)\Psi-2(1-u) \Psi' \,, 
\nonumber \\
& {\cal H}_{rr}=\frac{\ell^2}{r_h^2}\left(3\Psi- 2 u \Psi' \right) \,, 
\nonumber \\
& {\cal H}_{\varphi \varphi}=\left(\frac{3}{u}-1\right)\Psi-2(1-u) \Psi'  \,, 
\end{align}
and the exact solution of $\Psi$ satisfying the boundary condition (\ref{seisoku_con}) is given by 
\begin{align}
\Psi(u)=u^{(3-p)/2} F \left(\frac{1-p}{2},\,  \frac{3- p}{2},\,2 \,;1-u \right) \,. 
%
\end{align}

Figure~\ref{static_semiclassical_BH} shows 
the profile of ${\cal H}_{vv}$ interpolating the horizon to the AdS boundary in the $\hat{m}^2=-4/5$ case. 
This is proportional to the energy density detected by an observer along the timelike Killing orbit of $\p_v$. The unit tangent 
vector along the orbit is given by $V^\mu=(\p_v)^\mu/\sqrt{f}$. 
From eqs.~(\ref{stress_energy_tensor_formula}) and (\ref{relation_a0_a3}), 
we obtain 
\begin{align}
\label{null_energy}
\Exp{\cal{T}_{\mu\nu}}V^\mu V^\nu 
 \sim -\epsilon \hat{m}^2{\cal H}_{vv} \,.
\end{align}
So, when $\epsilon>0$, the energy density is everywhere positive outside the black hole horizon.  

Figure~\ref{stress_energy_fig} shows the stress energy tensor, $\langle{{\cal{T}}_t}^t \rangle$, $\langle{{\cal{T}}_r}^r \rangle$, $\langle{\cal T}_{\varphi}^\varphi \rangle$, and $f {\cal H}_{rr}$ 
are shown 
for $\hat{m}^2=-4/5$. 
By coordinate transformation, 
$v=t+\int dr f(r)^{-1}$, we obtain usual diagonal metric from the Eddington-Finkelstein form~(\ref{v_r_coor}). 
By eq.~(\ref{stress_energy_tensor_formula}), the stress-energy tensor $\langle \cal{T}_{\mu\nu} \rangle$ is proportional to the perturbed metric $h_{\mu\nu}$, given by  
\begin{align} 
\label{perturb_static_diagonal}
{h_t}^t=-\epsilon {\cal H}_{vv}, \qquad {h_r}^r=-\epsilon({\cal H}_{vv}-f{\cal H}_{rr}), \qquad 
 {h_\varphi}^\varphi=\epsilon(2{\cal H}_{vv}-f{\cal H}_{rr}) \,.
\end{align}
Near the horizon $r=r_h$, $f{\cal H}_{rr}$ is very small, and then  
we approximately write the form of the stress-energy tensor as 
\begin{align}
\Exp{{{\cal T}_t}^t} \simeq \Exp{{ {\cal T}_r}^r} \simeq -\frac{\Exp{{\cal{T}_\varphi}^\varphi} }{2} \,.
\end{align}
This form is very similar to the one of a free conformal scalar field~\cite{Steif:1993zv,Shiraishi:1993nu,Shiraishi:1993qnr,Lifschytz:1993eb} and for a holographic 
CFT dual to a four-dimensional bulk~\cite{Hubeny:2009rc}, 
\begin{align}
\label{stress-energy_conformal}
\Exp{{\cal{T}_\nu}^\mu}\sim \frac{1}{r^3}\mbox{diag}(1, 1, -2) \,.
\end{align} 
On the other hand, we find that $f{\cal H}_{rr}$ grows toward the AdS boundary and $f{\cal H}_{rr}\simeq {\cal H}_{vv}$ 
near $u=0$ by eq.~(\ref{static_HvrKHrr}). This means that $\Exp{{\cal{T}}_r^r}$ quickly approaches zero, compared with the other 
components, $\Exp{ {\cal{T}}_t{}^t } $ and $\Exp{ \cal{T}_{\varphi}{}^\varphi } $ near the AdS boundary, being quite different from the form~(\ref{stress-energy_conformal}).

According to the AdS/CFT dictionary, the states of the boundary field theory is represented by the various geometries of the bulk 
solutions. In our holographic setting, the bulk black hole connects two boundary black holes located on the north and south 
poles, while in the holographic model of the BTZ black hole~\cite{Hubeny:2009rc}, the bulk geometry and the horizon 
cap off smoothly on a bubble, and the boundary black hole is isolated from the other bulk horizon. The former is usually referred to 
as {\it black funnels}, while the latter is referred to as {\it black droplets}~\cite{Hubeny:2009ru}. See figure~\ref{geom:fnldrplt} below. 
So, there are two possibilities. One possibility is that the difference of the boundary stress-energy tensor reflects the difference of the 
bulk geometry. Another possibility is that our static solution is unstable against generic perturbations.

\begin{figure}[h]
  \begin{center}
  \includegraphics[width=10cm]{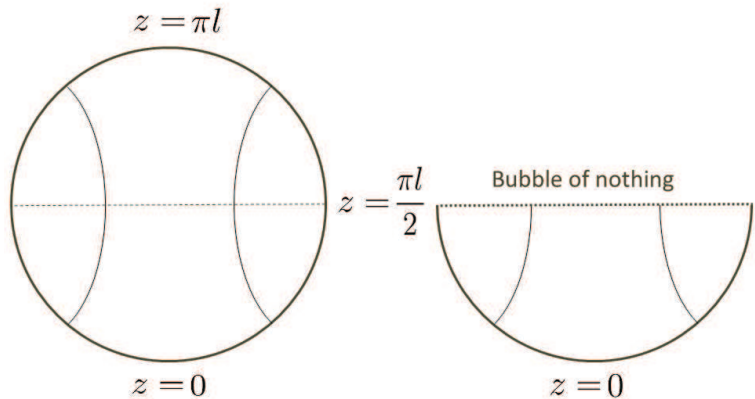}
  \end{center}
 \vspace{-0.5cm} 
  \caption{The bulk geometries of a black funnel (left) and a black droplet (right). The coordinate system similar to (\ref{metric_BTZ_BS}) is taken so 
that the thick (semi-)circles, $z=0, \pi l$, denote the AdS boundary, the thin curves inside the bulk correspond to the bulk horizon, which connects two AdS boundaries in the black funnel case (left), and which ends at the bubble of nothing in the black droplet case (right). } 
 \label{geom:fnldrplt}
\end{figure} 


\subsubsection{Time-dependent perturbations} 
Next, we consider time-dependent perturbation of the BTZ solution satisfying eq.~(\ref{linearized_semi}) to show that the BTZ solution is unstable when the three-dimensional gravitational constant is larger than $G_{3, {\rm min}}$ in eq.~(\ref{lower_bound_G3}). As shown below, the solution of eq.~(\ref{Basic_eq_H_munu}) is analytically obtained, under our metric ansatz. 

In terms of the coordinate variable $u=r_h^2/r^2$ introduced just below eq.~(\ref{master_static}), the background BTZ metric given in eq.~(\ref{metric_BTZ_BS}) is cast into the form,  
\begin{align}
\label{BTZ_tu} 
\bar{ds}_3^2=-\frac{r_h^2}{\ell^2u}g(u)dt^2+\frac{\ell^2du^2}{4u^2g(u)}+\frac{r_h^2}{u}d\varphi^2 \,, 
\quad g(u)=1-u \,. 
\end{align}  
Introducing four functions $T(u)$, $U(u)$, $S(u)$, and $R(u)$, we make the metric ansatz 
of the perturbation as 
\begin{eqnarray} 
\label{perturbation_t_depend_BTZ}
 ds_3^2 &=& -\frac{r_h^2g(u)}{\ell^2u}(1+\epsilon T(u)e^{-i\omega t})dt^2+
\frac{\ell^2}{4u^2g(u)}(1+\epsilon U(u)e^{-i\omega t})du^2 
\nonumber \\
 &+&\frac{\epsilon r_h}{2u^2g(u)}S(u)e^{-i\omega t}dudt +\frac{r_h^2}{u}(1+\epsilon e^{-i\omega t}R(u))d\varphi^2 \,. 
\end{eqnarray}
  
From eq.~(\ref{transverse-traceless}), we obtain the following three constraint equations 
\begin{align} 
\label{time_depend_traceless_con}
R+U+T=0 \,,  
\end{align}
\begin{align} 
& \left(u\frac{d}{du}-2\right)S-i\hat{\omega} uT=0 \,, 
\label{time_depend_con:ST}
\end{align}
\begin{align} 
& \frac{i\hat{\omega}S}{2g}+ug'(U-T)+g(R+T-2U+2uU')=0 \,,
\label{time_depend_transverse_con}
\end{align}
where here and in the following $\hat{\omega}:=\ell^2\omega/r_h$.
By eq.~(\ref{time_depend_traceless_con}), $R$ can be eliminated from eqs.~(\ref{time_depend_transverse_con}) and 
(\ref{Basic_eq_H_munu}), and then, the following constraint equation
\begin{align}
\label{constr_time_depend}
\{2\hat{\omega}(1+\hat{m}^2)g-2\hat{\omega}(1+\hat{\omega}^2)u\}U-i(\hat{m}^2-\hat{\omega}^2)S
-2u\hat{\omega}(1+\hat{\omega}^2)T=0 \,, 
\end{align}  
is derived. Eliminating $T$ from eqs.~(\ref{time_depend_con:ST}) and (\ref{time_depend_transverse_con}) by using eq.~(\ref{constr_time_depend}), we obtain two coupled differential equations,  
\begin{align} 
\label{Eq_UV}
& \left(u\frac{d}{du}-\frac{\hat{\omega}^2}{2g}-2\right)Z+i\hat{\omega}(1+\hat{\omega}^2)U=0 \,,  
\nonumber \\
& \left(u\frac{d}{du}+\frac{\hat{\omega}^2-2}{2g}\right)U
+\frac{1}{4g^2}\left(i\hat{\omega}+\frac{\hat{m}^2-\hat{\omega}^2}{i\hat{\omega}(1+\hat{\omega}^2)}g\right)Z=0 \,, 
\nonumber \\
& Z:=S+2i\hat{\omega}gU \,. 
\end{align}
Then, it is easy to derive the following master equation by eliminating $U$ as 
\begin{align} 
\label{master_Z}
Z''-\left(\frac{2}{u}+\frac{1}{g}\right)Z'+\frac{1}{4u^2g^2}\left[u\hat{\omega}^2+g(8-\hat{m}^2)  \right]Z=0 \,.
\end{align}

The solution is given by the hypergeometric function as 
\begin{align} 
\label{choukika}
Z_\pm=(1-u)^{-{i\hat{\omega}}/{2}}u^{(3\pm p)/2}
F\left(-\frac{1\mp p+i\hat{\omega}}{2},\, \frac{3\pm p-i\hat{\omega}}{2} ,\,1\pm p ;  \,u   \right) \,. 
\end{align}
By imposing Dirichlet boundary condition at $u=0$, we choose $Z_+$ solution only. Near the horizon, $u=1$, we also impose 
the ingoing boundary condition, $Z_+\sim (1-u)^{-i\hat{\omega}/2}$. By using Gaussian transform formula, the 
ingoing boundary condition becomes
\begin{align}
\hat{\omega} = i (1-p-2n) \,, 
\quad (n=0,1,2,,\cdots) \,.
\end{align} 
This gives an unstable mode in the range of $-1<\hat{m}^2<-3/4$ by setting the lowest mode, $n=0$ as 
\begin{align} 
\label{unstable_mode}
\hat{\omega}=i(1-p)=i(1-\sqrt{1+\hat{m}^2}) \,.
\end{align}
Substituting eq.~(\ref{unstable_mode}) into eq.~(\ref{choukika}), $Z_+$ reduces to a simple form:
\begin{align}
\label{exact_Z}
Z_+=(1-u)^{-{i\hat{\omega}}/{2}}u^{2+  {i\hat{\omega}}/{2}} \,. 
\end{align}
The existence of the unstable mode~(\ref{unstable_mode}) means that the BTZ black hole solution is unstable 
against the quantum field perturbation via the semiclassical Einstein equations~(\ref{linearized_semi}) when $G_3>G_{3, {\rm min}}$. %

One might wonder if the perturbed metric is regular on the horizon because both of the functions $T$ and $U$ diverge 
near the horizon, $u=1$ as
\begin{align}
T,\,U\sim (1-u)^{-1-i\hat{\omega}/2} \,. 
\end{align}
To remove this apparent singularity, let us use the advanced null coordinate~${v}$, which is now given by  
\begin{align}
\label{null_coordinate_v} 
d {v}= dt - \dfrac{\ell^2}{2r_h\sqrt{u}g}du \,. 
%
\end{align} 
Then, the metric (\ref{perturbation_t_depend_BTZ}) becomes  
\begin{align}
\label{metric:advEFtype}
& ds_3^2 = -\frac{r_h^2 g}{\ell^2u}\left(1+ \epsilon e^{-i\omega t} T \right)d {v}^2 
               - \frac{r_h}{u^{3/2}} \left\{ 1+\frac{\epsilon}{2}\left( Q- e^{-i\omega t}R \right) \right\} d{v} du 
              -\frac{\ell^2}{4u^2g}\epsilon Q du^2 
\nonumber \\
& \quad \qquad {} + \frac{r_h^2}{u}\left(1 + \epsilon e^{-i\omega t} R \right) d\varphi^2 \,.   
\end{align}
In particular near the horizon $u \rightarrow 1$, the above metric takes the regular form,  
\begin{align}
\label{near_horizon_metric}
& ds_3^2\simeq -\frac{r_h^2}{\ell^2u}\left(g+ \frac{\epsilon}{2} \frac{ e^{-i\omega {v} }}{1-i\hat{\omega}} \right)d {v}^2
-\frac{r_h}{u^{3/2} }\left(1+ \frac{\epsilon}{4} \frac{ e^{-i\omega {v} }}{1-i\hat{\omega}} \right) d {v} du 
 + \frac{\ell^2}{4u^2} \frac{\epsilon}{8} \frac{e^{-i\omega {v}} }{1- i\hat{\omega}} du^2
\nonumber \\ 
&
 \quad \quad +\frac{r_h^2}{u}\left(1- \frac{\epsilon}{2} \frac{e^{-i\omega {v} } }{1-i\hat{\omega} }   
                    \right)d\varphi^2 \,,  
\end{align}
where we have used the fact that $R$ is obtained from eq.~(\ref{time_depend_traceless_con}) and near the horizon $Q \propto (1-u)  \to 0$.

\subsection{Boundary dynamics and instability}
It is interesting to investigate how the boundary black hole evolves according to the semiclassical Einstein equations~(\ref{linearized_semi}).  
Unfortunately it is in general difficult to determine the location of the event horizon within the framework of the perturbation, as one needs to know the whole evolution of the geometry to determine it. Nevertheless, we can still obtain some insights for this problem by inspecting 
the behavior of energy flux across the apparent horizon within the perturbation framework. For this purpose, let us examine 
the metric $ds_3^2= {\cal G}_{\mu \nu}dx^\mu dx^\nu$ given by the near horizon expression~(\ref{near_horizon_metric}). 
Suppose the apparent horizon is located at $u_{\rm AH}= 1+ \epsilon \zeta e^{-i \omega {v}}$ with the out-going null tangent 
$l^\mu := (\partial_{{v}})^\mu + \epsilon \chi e^{-i \omega {v}} (\partial_u)^\mu$, where $\zeta$ and $\chi$ are some parameters. 
Then, it must follow that on $u=u_{\rm AH}$, $l^\mu l_\mu =0 $ and $l^\mu \partial_\mu {\cal G}_{\varphi \varphi}=0$. 
From these requirements, by using the concrete expressions of ${\cal G}_{\mu \nu}$ given in eq.~(\ref{near_horizon_metric}), 
we find 
\begin{align}
 \zeta = \frac{1+ i \hat{\omega}}{ 2(1-i\hat{\omega})} \,, \quad \chi = \frac{ i \kappa \hat{\omega}}{2(1-i\hat{\omega})} \,, 
\end{align} 
where here and hereafter $\kappa$ denotes the surface gravity of the background BTZ black hole, i.e., $\kappa := r_h /\ell^2$. Therefore we obtain 
\begin{align} 
\label{u:AH}
u_{\rm AH}=1+ \frac{\epsilon}{2} \frac{ p}{(2-p)}e^{\kappa (1-p) {v}} + O(\epsilon^2) \,, 
\end{align}
with $0<p=\sqrt{1+\hat{m}^2}<1$.  
The horizon radius $r_{\rm AH}$, defined by $r_{\rm AH}:= {\cal G}_{\varphi \varphi}^{1/2}(u=u_{\rm AH})$, 
can be evaluated, via eqs.~(\ref{near_horizon_metric}) and (\ref{u:AH}), as 
\begin{align} 
 \frac{r_{\rm AH}}{r_h}  \simeq 1 - \frac{\epsilon}{4} \frac{ 1+p }{2-p}e^{\kappa (1-p) {v}} \,. 
\end{align} 

When $\epsilon>0$, 
$r_{\rm AH}$ decreases along the direction of $l^\mu$, implying that the black hole evaporates due to quantum energy flux 
across the horizon. Indeed, the energy flux ${\cal F} :=\Exp{{\cal T}_{vv}}$ at $O(\epsilon)$ is evaluated by eqs.~(\ref{stress_energy_tensor_formula}) and (\ref{relation_a0_a3}). For the unstable mode satisfying (\ref{unstable_mode}), 
we find 
\begin{align}
\label{null_energy_flux} 
 8\pi G_3 {\cal F} \sim - \frac{\hat{m}^2}{2 \ell^2} \delta g_{(0) {v} {v}}(u=1) 
                       \sim - \frac{\epsilon }{4} \kappa^2\frac{ 1-p^2}{2-p} e^{\kappa (1-p) {v}} 
                        =       \frac{\kappa }{r_h} \: \frac{dr_{\rm AH}}{d {v}} \,,  
\end{align}
where we have used eq.~(\ref{unstable_mode}) in the last equality. 
The energy flux is negative for $\epsilon>0$, since we have a semiclassical solution only when $-1<\hat{m}^2<-3/4$, 
as shown in the previous section.

At $r=0~(u=\infty)$ spacelike singularity inside the BTZ black hole, the stress-energy tensor diverges as 
\begin{align}
\label{T^2_singularity}
\Exp{{\cal T}_{\mu\nu}}\Exp{{\cal T}^{\mu\nu}}\sim \epsilon^2 H_{\mu\nu}H^{\mu\nu}\sim \epsilon^2 e^{{2\kappa (1-p)t}}
(u-1)^{1-p}u^{1+p}\sim u^2\to \infty \,,   
\end{align}
which is a gauge invariant quantity, up to $O(\epsilon^2)$ because the background stress-energy tensor is zero. This suggests 
that the perturbed BTZ black hole has a curvature singularity at $u=\infty$ because the perturbed 
Einstein tensor $\delta \tEin_{(0)\mu\nu}$ of the boundary metric $g_{(0)\mu\nu}$ diverges at $O(\epsilon)$ as 
\begin{align}
\left(\tEin_{(0)\mu\nu}-\frac{g_{(0)\mu\nu}}{\ell^2}\right)\left(\tEin^{(0)\mu\nu}-\frac{g^{(0)\mu\nu}}{\ell^2}\right)=
\delta \tEin_{(0)\mu\nu}\delta \tEin^{(0)\mu\nu}  \to \infty \,, 
\end{align}
through the semiclassical Einstein equations~(\ref{linearized_semi}). So, the {\em quantum corrected} BTZ black hole is singular 
at $u=\infty$, although the unperturbed one has no curvature singularity there.   
The boundary singularity is also a bulk curvature singularity at $u=\infty$. This is observed by 
the calculation of $C_{KLMN}C^ {KLMN}$ at $O(\epsilon^2)$ as 
\begin{align}
\label{Weyl_invariant}
C_{KLMN}C^{KLMN}=R_{KLMN}R^{KLMN}-\frac{24}{L^4}\sim 
\epsilon^2\left(\frac{\ell}{L^3}\xi(z)\sin^3\frac{z}{\ell}\right)^2e^{{2\kappa (1-p) t}}u^4 \,, 
\end{align}
near $u=\infty$, which is also gauge invariant, up to $O(\epsilon^2)$ by $C_{KLMN}=O(\epsilon)$. 
The appearance of the bulk curvature singularity implies that {\it generic inhomogeneous} AdS$_4$ black string contains a curvature singularity 
inside the black hole, even though the background BTZ black string has no real curvature singularity. 

It is interesting to explore what is the endpoint of the black hole evaporation. If the boundary black hole completely evaporates until the radius reaches the Planck scale, a naked singularity would appear on the boundary. From the bulk point of view, the initial black string solution in eq.~(\ref{metric_BTZ_BS}) becomes inhomogeneous under the evolution, and the horizon pinch-off occurs 
at the AdS boundary, via the complete evaporation of the boundary black hole. 
This is analogous to the Gregory-Laflamme instability of the black string solution~\cite{Gregory:1993vy}.   
In that case, the central curvature singularity would be seen by a bulk observer outside 
the AdS black hole, implying the violation of the weak cosmic censorship in asymptotically AdS$_4$ spacetime. 
This picture is consistent with the recently study in the Einstein-Maxwell~(EM) system~\cite{Horowitz:2016ezu}. 
In our semiclassical model, the naked singularity would be caused by the boundary metric determined by the 
semiclassical Einstein equations~(\ref{linearized_semi}). 

\section{Summary and discussions} 
\label{sec:4}
We have investigated semiclassical Einstein equations in the framework of the AdS/CFT correspondence. 
We considered perturbations of the pure AdS$_{d+1}$ spacetime and derived a mixed boundary condition, 
which corresponds to the semiclassical Einstein equations in which the ratio between the faster and the slower fall-off modes is determined by the dimensionless parameter $\gamma_d$. In particular, we have analytically solved 
the perturbation equations in the $d=3$ case and found that the BTZ black hole with vanishing expectation value of the stress-energy tensor is {\em semiclassically unstable}, provided that the parameter $\gamma_3$ exceeds the critical value (i.e., $\gamma_3>1$), or equivalently the $3$-dimensional gravitational constant $G_3$ is above the critical value (i.e., $G_3 > G_{3, {\rm min}}$). The occurrence of this instability can be interpreted as follows. When the gravitational constant is small enough, the mixed boundary condition reduces to the ordinary Dirichlet boundary condition at the AdS boundary under which the BTZ black hole is stable against any perturbations~\cite{Liu:2008ds}. However, as one increases the gravitational constant, the backreaction of the quantum stress-energy tensor becomes important and it leads to a new instability caused by the quantum effect\footnote{
It would be interesting to study whether a similar type of instabilities could also occur for the gauge field case.  
Although our preliminary calculations indicate no such instabilities to occur, we need further study to reach a definite conclusion. 
}.

Strictly speaking, we have examined the perturbations in the geometry of the pure AdS$_3$ expressed in terms of the BTZ coordinates, rather than the decent BTZ black hole, which is constructed by making certain identification of spacetime points of the pure AdS$_3$, thus having a topologically non-trivial geometry. Accordingly the quantum field resides 
on our boundary AdS$_3$ with BTZ coordinates (or the covering space of a BTZ black hole) is not in the Hartle-Hawking state with finite temperature, 
but is rather in the conformal vacuum with zero-temperature. (For this reason, we have referred to our boundary setting as the BTZ black hole with vanishing expectation value of 
the stress-energy tensor.) The instability found in this paper can therefore be viewed as an instability (or phase transition) of AdS$_3$ itself due to the effects of strongly coupled quantum fields. 
It is important to check whether this type of instability of pure AdS with quantum fields in conformal vacuum state can also occur in higher dimensions $d>3$. In this regard, it is also interesting to study the role of trace anomalies when $d$ is even.

There remains an open issue whether the bulk spacetime dynamically evolves in a manner consistent with the semiclassical Einstein equations at the AdS boundary, since there are many gravitational corner conditions~\cite{Horowitz:2019dym}. 
Although further careful analysis is required beyond the linear perturbation, it would be interesting to explore what is the final state of the instability. 
One possibility is that the final state of the bulk spacetime is a non-uniform black funnel solution, connecting two boundary AdS black holes~\cite{Marolf:2019wkz,Emparan:2021ewh}.         
Another possibility may be that the time evolution does not settle down to any stationary solution, such as the static black funnel. When the boundary black hole evaporates completely, a naked singularity appears on the boundary, as argued in the previous section. When the boundary black hole horizon continues to expand, the bulk horizon of the black funnel gets thin by the outflow of null generators from the bulk into 
the AdS boundary. In either case, pinch-off of the bulk horizon would occur inside the bulk or on the boundary within a finite time, analogous 
to the Gregory-Laflamme instability~\cite{Gregory:1993vy}. Our analysis can be extended into the higher dimensional cases~($d>3$) by perturbing higher dimensional AdS$_{d+1}$ black string solutions. In higher dimensional case, there is another control parameter describing the black hole radius, in addition to the dimensionless parameter $\gamma_d$. As the semiclassical effect renders the BTZ black hole unstable, 
it is natural to speculate that the parameter region in which the higher dimensional AdS black string solution is stable becomes smaller than the one analyzed under the Dirichlet boundary condition~\cite{Marolf:2019wkz}. 
The complete analysis would appear in the near future~\cite{IshibashiMaedaOkamura}. 
   
Many other interesting directions can be explored by applying the holographic semiclassical method. One of them is the issue of the black hole evaporation. 
The backreaction problem of the Hawking radiation has been investigated in the two-dimensional dilaton model~\cite{Callan:1992rs}, but only few 
have so far been examined in higher-dimensional models. 
Although the semiclassical approximation is violated at the final stage of the evaporation, it would be valuable to investigate how the black hole evolves with evaporation. 
Another direction is to test the self-consistent averaged null energy condition~(ANEC)~\cite{Graham:2007va} in our holographic semiclassical approach. 
The ``self-consistent'' means that the expectation value of the quantum stress-energy tensor~(in addition with other classical 
stress-energy tensor) satisfies the semiclassical Einstein equations. 
This avoids violation of the ANEC observed in conformally coupled scalar field in a conformally flat spacetime~\cite{Urban:2009yt}, or a strongly coupled field in a holographic setting~\cite{Ishibashi:2019nby} in the absence of the self-consistent condition. If the self-consistent ANEC holds in general spacetime, we can also establish various important theorems such as the singularity theorem or the topological censorship, incorporating the quantum effect semiclassically.

One way of viewing the holographic semiclassical Einstein equations might be in terms of AdS/BCFT~\cite{Takayanagi:2011zk} (see also \cite{Izumi:2022opi} 
and references therein).  
In the AdS/BCFT setting, one includes an end-of-the-world (EOW) brane whose end at the AdS boundary corresponds to the boundary of BCFT, so that the gravity dual of BCFT is given by the region of AdS bulk surrounded by the AdS boundary and the EOW brane. 
The bulk metric is required to satisfy the Neumann boundary condition at the EOW brane, rather than the Dirichlet boundary condition. This in turn implies that the metric on the EOW brane becomes dynamical, satisfying the braneworld Einstein equations~\cite{Shiromizu:1999wj}, and thus correspond, as a gravity dual, to the boundary degrees of freedom of the BCFT. 
One may expect that by taking the limit toward the AdS conformal boundary, the braneworld Einstein equations would turn to play a role of the boundary semiclassical Einstein equations. It would be interesting to clarify the possibility of such relations between the braneworld gravity and holographic semiclassical Einstein equations.

\bigskip
\goodbreak
\centerline{\bf Acknowledgments}
\noindent

This work was supported in part by JSPS KAKENHI Grant No. 15K05092, 20K03938, 20K03975, 17K05451 (A.I.), 20K03975 (K.M.), and 17K05427(T.O.), 
and also supported by MEXT KAKENHI Grant-in-Aid for Transformative Research Areas A Extreme Universe No. 21H05182, 21H05186 (A.I. and K.M.).


%
\appendix
\section{Derivation of the permittivity and permeability from the holographic semiclassical Maxwell equations}
\label{sec:Maxwell_exmpl}

As an example of our holographic semiclassical Maxwell equations~(\ref{eq:EOM-holo_semiC-Maxwell_exmpl-all}), 
we derive the permittivity and permeability~\cite{Natsuume:2022kic} at finite temperature. 
Let us consider the semiclassical problem (\ref{eq:holo_semiC-action-Maxwell_exmpl}) which couples to 
$\pExp{\calJ_\mu}$ in dual field theory at finite temperature. 
For this purpose, we consider as our bulk gravity Schwarzschild-AdS black hole (SAdS$_4$), 
expressed in the Poincar\'e like coordinates
$X^M = (x^\mu\,;\, z) = (t, x, y\,;\, z)$ as 
\begin{align}
  & ds_{4}^2
  = \left( \frac{L}{z} \right)^2\,
  \left( - f(z)\, dt^2 + dx^2 + dy^2 + \frac{dz^2}{{f(z)} }
  \right) \,, 
& & f(z) := 1 - (\calT\, z)^3 
  ~, 
\label{A-eq:asympt-bulk_metric}
\end{align}
where the AdS conformal boundary is at $z = 0$ and where $\calT := 4\, \pi\, T_{\tBH}/3$ with 
$T_\tBH$ being the Hawking temperature. 


In order to solve eqs.~(\ref{eq:EOM-holo_semiC-Maxwell_exmpl-all}), 
let us assume $A_z = 0$ and take the Fourier transformation along the boundary as  
\begin{align}
  & A_\mu(x, z)
  = \int \frac{d^3k}{ (2\, \pi)^{3/2} }~e^{i k \cdot x}\, \tilA_\mu(k, z)
  ~,
& & k_\mu = (- \omega,\, q,\, 0)
  ~.
\label{eq:FT--Maxwell_exmpl}
\end{align}
Then, by taking expansion with respect to $\mfw := \omega/\calT \ll 1$, $\mfq := q/\calT \ll 1$ one can find the solutions. 
For simplicity, consider the stationary configuration $\omega = 0$, in which eqs.~(\ref{eq:EOM-holo_semiC-Maxwell_exmpl-bulk}) reduce to
%


\begin{subequations}
\label{A-eq:FT_EOM-holo_semiC-Maxwell_exmpl-bulk-all}
\begin{align}
  & 0
  = \left( - f\, \partial^2_z + q^2 \right)\, \tilA_t
  ~,
\label{A-eq:FT_EOM-holo_semiC-Maxwell_exmpl-bulk-t} \\
  & 0
  = \partial_z \tilA_x
  ~,
\label{A-eq:FT_EOM-holo_semiC-Maxwell_exmpl-bulk-x} \\
  & 0
  = \left\{ \partial_z (f\, \partial_z) - q^2 \right\}\, \tilA_{y}
  ~.
\label{A-eq:FT_EOM-holo_semiC-Maxwell_exmpl-bulk-y}
\end{align}
\end{subequations}
%
The boundary conditions at the horizon $z = z_H := 1/\calT$ are given by 
\begin{align}
  & \tilA_t(q, z_H) = 0~,
& & \big|\, \tilA_x(q, z_H)\, \big| < \infty~,
& & \big|\, \tilA_y(q, z_H)\, \big| < \infty
  ~.
\label{A-eq:bc-holo_semiC-Maxwell_exmpl-bulk-all}
\end{align}
%
Expanding eqs.~(\ref{A-eq:FT_EOM-holo_semiC-Maxwell_exmpl-bulk-all}),
(\ref{A-eq:bc-holo_semiC-Maxwell_exmpl-bulk-all}) with respect to $\mfq^2$, we 
find the following solutions:
\begin{subequations}
\label{A-eq:sol-phi_L-SAdS_4-FT-hydro_exp-all}
\begin{align}
  & \tilA_t(q, z)
  = \frac{ 1 - \calT\, z + \mfq^2\, I(\calT z) }{ 1 + \mfq^2\, I(0) }\,
    \tcA_t(q)
  + O(\mfq^4)
  ~,
\label{A-eq:FT_sol-holo_semiC-Maxwell_exmpl-bulk-t} \\
  & \tilA_x(q, z)
  = \tcA_x(q)
  ~,
\label{A-eq:FT_sol-holo_semiC-Maxwell_exmpl-bulk-x} \\
  & \tilA_y(q, z)
  = \frac{ 1 - \mfq^2\, I'(\calT z) }{ 1 - \mfq^2\, I'(0) }\, \tcA_y(q)
  + O(\mfq^4)
  ~,
\label{A-eq:FT_sol-holo_semiC-Maxwell_exmpl-bulk-y}
\intertext{where}
  & I(\eta)
  := \int^1_{\eta} d\eta' \int^1_{\eta'}
    \frac{ d\eta'' }{ 1 + \eta'' + \eta''^2 }
  ~.
\end{align}
\end{subequations}
Then, from eq.~(\ref{eq:dictionary-current}), we obtain the R-current as
\begin{subequations}
\label{eq:R-current-stationary-all}
\begin{align}
  & \pExp{\calJ^t}
  = \frac{\calT}{g^2}\, \left\{ 1
    + \frac{\mfq^2}{2}\, \left( \frac{\pi}{\sqrt{3}} - \ln 3 \right)
  \right\}\, \tcA_t(q)
  + O(\mfq^4)
  ~,
\label{eq:R-current-stationary-t} \\
  & \pExp{\calJ^x}
  = 0
  ~,
\label{eq:R-current-stationary-x} \\
  & \pExp{\calJ^y}
  = - \frac{\calT}{ g^2 }\, \mfq^2\, \tcA_y(q)
  + O(\mfq^4)
  ~.
\label{eq:R-current-stationary-y}
\end{align}
\end{subequations}
Here we have used the following:
\begin{align*}
& I(0)
= \frac{ 9\, \ln 3 - \sqrt{3}~\pi }{18}
~,
&& I'(0)
= - \int^1_{0} \frac{d\eta}{1 + \eta + \eta^2}
= - \frac{\pi}{ 3\, \sqrt{3} }
~,
& & I''(0) = 1
 ~.
\end{align*}
%
Using the above formulas, we can express eqs.~(\ref{eq:EOM-holo_semiC-Maxwell_exmpl-bdy}) as follows: 
\begin{subequations}
\label{eq:FT_EOM-holo_semiC-Maxwell_exmpl-bdy-all}
\begin{align}
  & 0
  = - q^2\, \tcA_{t}
  - \frac{\calT\, e^2}{g^2}\, \left\{ 1
    + \frac{\mfq^2}{2}\, \left( \frac{\pi}{\sqrt{3}} - \ln 3 \right)
  \right\}\, \tcA_t
  - e^2\, \calJ_\tExt^t
  \propto \left( q^2 + \frac{\calT}{g^2\, \epsilon} \right)\, \tcA_t
  - \frac{ \calJ_\tExt^t }{\epsilon}
  ~,
\label{eq:FT_EOM-holo_semiC-Maxwell_exmpl-bdy-t} \\
  & 0
  = - e^2\, \calJ_\tExt^x
  ~,
\label{eq:FT_EOM-holo_semiC-Maxwell_exmpl-bdy-x} \\
  & 0
  = q^2\, \tcA_{y}
  + \frac{\calT\, e^2}{ g^2 }\, \mfq^2\, \tcA_y(q)
  - e^2\, \calJ_\tExt^y
  \propto q^2\, \tcA_y(q) - \mu\, \calJ_\tExt^y
  ~.
\label{eq:FT_EOM-holo_semiC-Maxwell_exmpl-bdy-y}
\end{align}
\end{subequations}
where $\epsilon$, $\mu$ are defined as
\begin{subequations}
\begin{align}
  & \epsilon
  := \frac{1}{e^2}\, \left( 1
    + \frac{ \pi - \sqrt{3}~\ln 3 }{ 2\, \sqrt{3} }\,
      \frac{e^2}{g^2\, \calT} \right)
  \sim \frac{1}{e^2}\, \left( 1 + 0.36 \times
      \frac{e^2}{g^2\, \calT} \right)
  ~,
\label{eq:def-permittivity} \\
  & \mu
  := \frac{e^2}{ 1 + e^2/(g^2\, \calT) }
  ~. 
\label{eq:def-permeability}
\end{align}
\end{subequations}
We note that the dual field theory with no current $\pExp{\calJ^\mu} = 0$ corresponds to the limit of $g \to \infty$ in 
eqs.~(\ref{eq:FT_EOM-holo_semiC-Maxwell_exmpl-bdy-t}) and (\ref{eq:FT_EOM-holo_semiC-Maxwell_exmpl-bdy-y}). 
Taking this limit, we can find that $\epsilon$, $\mu$ correspond, respectively, the permittivity and permeability\footnote{Note that in our convention, the permittivity of vacuum corresponds to $\epsilon_0 = 1/e^2$ 
and the permeability of vacuum to $\mu_0 = e^2$. }, 
and also that $\calT/(g^2\, \epsilon) =: 1/\lambda^{2}_D$ in eq.~(\ref{eq:FT_EOM-holo_semiC-Maxwell_exmpl-bdy-t}) 
represents the screening effect due to the dual field with Debye length $\lambda_D$. 


\section{The expression of $\pExp{\calT_{\mu\nu}}$}
\label{sec:calT}
The expectation value of the stress-energy tensor~(\ref{eq:energy_stress-calZ}) is given by~\cite{deHaro:2000vlm,Balasubramanian:1999re}:
\begin{align}
   \pExp{\calT^{\mu\nu}}
  &= \frac{2}{\sqrt{- \bdyg}}\,
    \frac{ \delta S^\tos_\tbulk[\, \bdyg\,] }{\delta \bdyg_{\mu\nu}}
  = \lim_{z \to 0} \frac{1}{ \Omega^{d+2}(z) }\, \frac{2}{\sqrt{- \indg}}\,
    \frac{ \delta S^\tos_\tbulk[\, \indg\, ] }{\delta \indg_{\mu\nu}}
\nonumber \\
  &= \lim_{z \to 0} {\frac{1}{ \Omega^{d+2}(z) }}\,
  \left( \frac{ - K^{\mu\nu} + \indg^{\mu\nu}\, K }
              { {8\, \pi\, G_{d+1}} }
    + {\frac{2}{\sqrt{- \indg}}}\,
      \frac{\delta S_\tct}{\delta \indg_{\mu\nu}} \right)
\nonumber \\
  &
  = \lim_{z \to 0} \frac{ \Omega^{-(d+2)}(z) }{ 8\, \pi\, G_{d+1}\, L }\,
  \Bigg\{ - L\, K^{\mu\nu} + \indg^{\mu\nu}\, L\, K
    - (d - 1)\, \indg^{\mu\nu}
    + \frac{L^2}{d - 2}\, \tEin^{\mu\nu}[\, \indg\, ] + \cdots \Bigg\}
  ~, 
\label{A-eq:exp-calT-ori}
\end{align}
where the geometric quantities appeared here are given in Sec.~\ref{sec:formalism}, 
the conformal factor is assumed to fall-off as eq.~(\ref{eq:bulk_metric-cano_form}), and as for the counter-terms, only terms needed in $d \leq 3$ are written.   

Now we use the conformally rescaled metric and extrinsic curvature, ${\tilde g}_{\mu\nu} := \Omega^2\, {g}_{\mu\nu}$ and the extrinsic curvature $\tilde{K}_{\mu\nu} := - \partial_z \tilde{g}_{\mu\nu}/2$ given by eq.~(\ref{eq:def-barK}), which are regular at $z \to 0$. In terms of which, we can rewrite eq.~(\ref{A-eq:exp-calT-ori}) as  
\begin{subequations}
\label{A-eq:exp-calT_lower-all}
\begin{align}
  & \pExp{\calT_{\mu\nu}}
  = \lim_{z \to 0} \frac{ \Omega^{-d} }{ 8\, \pi\, G_{d+1}\, L }\, \bigg[\,
  - L\, \Omega\, \left( \tilde{K}_{\mu\nu} - {\tilde g}_{\mu\nu}\, \tilde{K} \right)
  + \frac{ L^2\, \Omega^2 }{d - 2}\, \tEin_{\mu\nu}[\, {\tilde g}\, ]
\nonumber \\
  &\hspace*{2.5truecm}
  - (d - 1)\, \left( 1 - L\, \Omega' \right)\, {\tilde g}_{\mu\nu}
  + \cdots\, \bigg]
\label{A-eq:exp-calT_lower} \\
  &\hspace*{1.15truecm}
  = \lim_{z \to 0} \frac{1}{ 8\, \pi\, G_{d+1}\, L }\,
  \Bigg[\, L^2\,
    \frac{ \tEin_{\mu\nu}[\, \bulkg\, ] + \Lambda_{d+1}\, \bulkg_{\mu\nu} }
         { (d - 2)\, \Omega^{d-2} }
  + L^2\, \frac{ 2\, \tilde{K} \tilde{K}_{\mu\nu}
    - \indcg_{\mu\nu}\, (\tilde{K}^{\rho\sigma} \tilde{K}_{\rho\sigma} + \tilde{K}^2) }
               { 2\, (d - 2)\, \Omega^{d-2} }
\nonumber \\
  &\hspace*{2.0truecm}
  - L\, {\tilde g}_{\nu\rho}\,
    \left( \frac{L\, \Omega}{d - 2}\, \pdiff{}{z} + 1 \right)\,
    \frac{ \tilde{K}_\mu{}^\rho - \delta_\mu{}^\rho\, \tilde{K}}{\Omega^{d-1}}
\nonumber \\
  &\hspace*{3.0truecm}
  + \frac{d - 1}{d - 2}\, \frac{ {\tilde g}_{\mu\nu} }{ \Omega^{d} }\,
    \left\{ L^2\, \Omega^2\, \left( \frac{\Omega'}{\Omega} \right)'
      + 1 - \frac{d - 2}{2}\, \left( 1 - L\, \Omega' \right)^2
    \right\}
  + \cdots
  \, \Bigg]
  ~.
\label{A-eq:exp-calT_lower-by_K}
\end{align}
\end{subequations}
Note that the indices are raised and lowered by $\bdyg_{\mu\nu}$ in the left-hand side, 
and by ${\tilde g}_{\mu\nu}$ in the right-hand side. Note also that the first term of the right-hand side if 
eq.~(\ref{A-eq:exp-calT_lower-by_K}) vanishes when there are no matter fields involved, due to 
the bulk Einstein equations~(\ref{eq:EOM-holo_semiC-Einstein_eqn-bulk}). The second-line of the right-hand side 
is more suitable to examine the limit $z\to 0$.

For asymptotically AdS bulk satisfying the bulk Einstein equations~(\ref{eq:EOM-holo_semiC-Einstein_eqn-bulk}), 
the asymptotic fall-off behavior of each term of eq.(\ref{A-eq:exp-calT_lower-by_K}) can be seen as follows. 
First note that according to \cite{Skenderis:2002wp}, the induced metric $\tilde{g}_{\mu\nu}$ behaves, at $z \to 0$, as: 
\begin{align}
  & \tilde{g}_{\mu\nu}
  \sim \bdyg_{\mu\nu}\, (1 + \sharp\, z^2 + \flat \, z^4 + \cdots)
  + \tilde{g}^{(f)}_{\mu\nu}\, z^d\, (1 + \cdots)
  ~.
\label{A-eq:indcg-fall_off-expectation}
\end{align}
In this expansion, the first term (slow fall-off mode) has a contribution to $\tilde{K}_{\mu\nu}$ at$O(\Omega)$, 
while the second term (fast fall-off mode) at $O(\Omega^{d-1})$, and thus,  
$\tilde{K}_{\mu\nu} \sim O(\Omega)\, \{\sharp + \flat O(z^2) \} \bdyg_{\mu\nu} + O(\Omega^{d-1})\, \indcg^{(f)}_{\mu\nu}$.
Hence, under the fall-off (\ref{A-eq:indcg-fall_off-expectation}), we can estimate the fall-off behavior of 
each term of eq.~(\ref{A-eq:exp-calT_lower-by_K}) as follows: 
\begin{itemize}
\setlength{\itemsep}{0.2truecm}
\item The quadratic terms of $\tilde{K}_{\mu \nu}$ (the second-term in the first line) in the right-hand side 
of eq.~(\ref{A-eq:exp-calT_lower-by_K}) is at $O(\indcK^2/\Omega^{d-2}) = O(\Omega^{4-d})$ and only the slow mode is relevant. 
For $d \le 3$, these terms have no contribution.  

\item The second-line have contributions at $O(z^{4-d})$ from the slow fall-off mode of ${\tilde g}_{\mu \nu}$ and at $O(1)$ from the fast fall-off, when the following holds, 
  \begin{align}
    & \frac{L\, \Omega}{z} = 1 + O(z^2)
    ~.
  \label{A-eq:Omega-fall_off-1}
  \end{align}
This is because, due to the effect of 
  \begin{align}
   & \frac{L\, \Omega}{d - 2}\, \pdiff{}{z} + 1
   = 1 - \frac{L\, \Omega}{z}
  + \frac{L\, \Omega}{z}\, \left( 1 - z^{2-d}\, \pdiff{}{ z^{2-d} } \right)
    ~,
  \end{align}
the contribution from the term $\sharp z^2$ in (\ref{A-eq:indcg-fall_off-expectation}) becomes at 
the same order of that from the term $\flat z^4$ in (\ref{A-eq:indcg-fall_off-expectation}). 
\item The third-line has contributions at $O(\Omega^{4-d})$, if the following holds: 
  \begin{align}
    & \sigma(z) := \frac{ 1 - L\, \Omega/z }{z^2}
    \hspace{0.3truecm}
    \text{is in $C^2$ at $z = 0$}
    ~.
  \label{A-eq:Omega-fall_off-2}
  \end{align}
  %
\end{itemize}

In the $d = 3$ case, if the condition~(\ref{A-eq:Omega-fall_off-2}) is satisfied, 
eq.~(\ref{A-eq:exp-calT_lower-by_K}) takes the following form, 
\begin{subequations}
\label{A-eq:exp-calT_lower-by_K-d3-all}
\begin{align}
  & \pExp{\calT_{\mu\nu}}_{d=3}
  = \lim_{z \to 0} \frac{1}{ 8\, \pi\, G_{4}\, L }~
    \tilde{g}_{\nu\rho}\, \left( L\, \Omega\, \pdiff{}{z} + 1 \right)\,
    \frac{ - L\, \tilde{K} _\mu{}^\rho + \delta_\mu{}^\rho\, L\, \tilde{K}  }{\Omega^2}
\label{A-eq:exp-calT_lower-by_K-d3-1} \\
  &\hspace*{1.8truecm}
  = \lim_{z \to 0} \frac{1}{ 8\, \pi\, G_{4}\, L }~
    \tilde{g}_{\nu\rho}\, \left( z\, \pdiff{}{z} + 1 \right)\, \frac{L^2}{z^2}\,
    \left( - L\, \tilde{K} \mu{}^\rho + \delta_\mu{}^\rho\, L\, \tilde{K}  \right)
  ~.
\label{A-eq:exp-calT_lower-by_K-d3-2}
\end{align}
\end{subequations}
In particular, we have $\tilde{K}_{\mu\nu} = 0$, whenever the metric $d \tilde{s}_4^2=G_{MN}dX^MdX^N$ can be expressed in terms of some $\Omega$ which satisfies condition~(\ref{A-eq:Omega-fall_off-2}) and metric $\tilde{g}_{\mu\nu}(x)$ which is independent of $z$, so that  
\begin{align} &
 d \tilde{s}_{4}^2
  = \Omega^{-2}(z)\,
  \left( dz^2 + \tilde{g}_{\mu\nu}(x)\, dx^\mu dx^\nu \right) 
  ~.
\label{A-eq:special_metric-d3}
\end{align}
Then, it immediately follows from eq.~(\ref{A-eq:exp-calT_lower-by_K-d3-all}) that $\pExp{\calT_{\mu\nu}}_{d=3} = 0$. 


\section{Formulas for the holographic semiclassical Einstein equations}
\label{sec:formula-holo-semiC_Ein_eq}
In this section, we provide some perturbation formulas of our holographic semiclassical Einstein 
equations with pure AdS$_{d+1}$ background (\ref{eq:bg_metric-all}), considered 
in Sec.~\ref{sec:formalism}. For the perturbed metric $G_{MN}=\bar{G}_{MN}+\delta G_{MN}$, we impose the gauge conditions $\delta G_{zM}=0$ so that nontrivial part of the metric becomes 
$\Omega^2 G_{\mu \nu}= \tilde{g}_{\mu \nu}(x,z)
= \bar{g}_{\mu \nu}(x) + h_{\mu \nu}(x,z)$. The indices of the perturbation variables $h_{\mu \nu}$ are raised and lowered by the background metric $\bar{g}_{\mu \nu}, \: \bar{g}^{\mu \nu}$. We denote the covariant derivative 
with respect to $\bar{g}_{\mu \nu}$ by $\bar{D}_\mu$ and introduce the projection operator 
\begin{equation}
\bar{P}_{\mu \nu} = \bar{D}_{(\mu} \bar{D}_{\nu)} - \dfrac{1}{d} \bar{g}_{\mu \nu} \bar{D}^2 \,.
\end{equation}
Then, the perturbation variables are decomposed as 
\begin{equation}
 h_{\mu \nu} = h_L \bar{g}_{\mu \nu} + \bar{P}_{\mu \nu} h_T^{(0)} + 2\bar{D}_{(\mu} h_{T\nu)}^{(1)}
 + h_{T\mu \nu}^{(2)} \,, 
\end{equation}
where $\bar{D}^\mu h^{(1)}_{T\, \mu} = 0$, $\bar{D}^\nu h^{(2)}_{T\, \mu\nu} = h^{(2)}_{T\, \mu}{}^\mu = 0$ 
are satisfied.

\subsection{Perturbation formulas for the bulk Einstein equations}
The bulk Einstein equation (\ref{eq:EOM-holo_semiC-Einstein_eqn-bulk}) is expressed 
as $E_{MN}=0$ in terms of the tensor defined by, 
\begin{align}
  & E_{MN}
  := R_{MN}[\, \bulkg\, ] + \frac{d}{L^2}\, \bulkg_{MN}
  ~.
\label{A-eq:def-E_MN}
\end{align}
For the case of the pure AdS$_{d+1}$ background (\ref{eq:pert_metric}), 
the linear perturbations of $E_{MN}$ are given, in terms of the gauge-invariant 
variable~(\ref{eq:gauge_inv-scalar_GW_L}), by the following formulas:
%
\begin{subequations}
\label{A-eq:delta-E_MN-rad_gauge-all}
\begin{align}
  & \delta E_{zz}
  = \left( \partial_{z} - \frac{\Omega'}{\Omega} \right)\, \partial_z h
  = \Omega\, \pdiff{}{z}\, \left( \frac{\partial_z h}{\Omega} \right)
  = d \times \Omega\,
  \pdiff{}{z}\, \left( \frac{\partial_z h_L}{\Omega} \right)
  ~,
\label{A-eq:delta-E_MN-rad_gauge-z^z} \\
  & \delta E_{z\mu}
  = - \partial_z \left( \bar{D}_\rho h_\mu{}^\rho
    - \bar{D}_\mu h \right)
\nonumber \\
  &\hspace*{0.9truecm}
  = (d - 1)\, \bar{D}_\mu \partial_z
    \left( \Phi + \frac{1}{\ell^2}\, h^{(0)}_T \right)
  - \left( \bar{D}^2 - \frac{d - 1}{\ell^2} \right)
    \partial_z h^{(1)}_{T\, \mu}
  ~,
\label{A-eq:delta-E_MN-rad_gauge-mu^z} \\
  & \delta E_{\mu\nu}
  = \frac{ \Omega^{d-1} }{- 2}\, \partial_{z}
    \left( \frac{ \partial_z h_{\mu\nu} }{ \Omega^{d-1} } \right)
  + \bar{D}_{(\mu} \bar{D}^{\rho} h_{\nu)\rho}
  - \frac{1}{2}\, \bar{D}^2 h_{\mu\nu}
  + \frac{ \bar{g}_{\mu\nu} }{2}\, \frac{\Omega'}{\Omega}\, \partial_z h
  - \frac{1}{2}\, \bar{D}_{\mu} \bar{D}_{\nu} h
\nonumber \\
  &\hspace*{1.8truecm}
  - \frac{1}{\ell^2}\,
    \left( h_{\mu\nu} - \bar{g}_{\mu\nu}\, h \right)
\nonumber \\
  &\hspace*{0.9truecm}
  = \bar{g}_{\mu\nu}\,
  \left[\,
    \frac{ \Omega }{- 2}\,
    \partial_{z} \left( \frac{ \partial_z h_L }{\Omega} \right)
  + (d - 1)\, \frac{\Omega'(z)}{\Omega}\, \partial_z h_L
  - \frac{d - 1}{d}\, \left( \bar{D}^2 - \frac{d}{\ell^2} \right)\, \Phi
  \, \right]
\nonumber \\
  &\hspace*{1.0truecm}
  - \frac{1}{2}\, \bar{P}_{\mu\nu}
  \left\{ \Omega^{d-1}\, \partial_{z}
      \left( \frac{ \partial_z h^{(0)}_T }{ \Omega^{d-1} } \right)
    + (d - 2)\, \Phi
  \right\}
  - \bar{D}_{(\mu}\, \left\{ \Omega^{d-1}\, \pdiff{}{z}\,
    \Bigg( \frac{ \partial_z h_{T\, \nu)}^{(1)} }{ \Omega^{d-1} } \Bigg)
    \right\}
\nonumber \\
  &\hspace*{1.5truecm}
  - \frac{1}{2}\, \left\{ \partial^2_z h^{(2)}_{T\, \mu\nu}
    - (d - 1)\, \frac{\Omega'(z)}{\Omega}\,
      \partial_z h^{(2)}_{T\, \mu\nu}
    + \left( \bar{D}^2 + \frac{2}{\ell^2} \right)\, h^{(2)}_{T\, \mu\nu}
  \right\}
  ~. 
\label{A-eq:delta-E_MN-rad_gauge-mu^nu}
\end{align}
\end{subequations}
%

\subsection{Perturbation formulas for the holographic semiclassical Einstein equations}
\label{App:C:pert} 

We write perturbation formulas for the holographic semiclassical Einstein equations~(\ref{eq:EOM-holo_semiC-Einstein_eqn-bdy}). The left-hand side of eq.~(\ref{eq:EOM-holo_semiC-Einstein_eqn-bdy}) is given by 
the limit $z \to 0$ of 
$ \tEin_{\mu\nu}[\, \tilde{g}\, ] + \Lambda_d\, \tilde{g}_{\mu\nu} $. 
Noting that $\Lambda_d = - (d - 1)\, (d - 2)/(2\, \ell^2)$, we find the following expression: 
\begin{align}
  \delta \left( \tEin_{\mu\nu}[\, \tilde{g}\, ]
  + \Lambda_d\, {\tilde{g}}_{\mu\nu} \right)
 &= \frac{ (d - 1)\, (d - 2) }{2\, d}\, \bar{g}_{\mu\nu}\,
 \left( \bar{D}^2 - \frac{d}{\ell^2} \right)\, \Phi
 - \frac{d - 2}{2\, d}\, \bar{P}_{\mu\nu}\, \Phi
 \nonumber \\
 &\hspace*{0.5truecm}
\nonumber \\
  &\hspace*{2.0truecm}
  - \frac{1}{2}\,
 \left( \bar{D}^2 + \frac{2}{\ell^2} \right)\, h_{T\, \mu\nu}^{(2)}
 ~.
\label{A-eq:del_semiC_Einstein_eqn-LHS}
\end{align}

Next we derive an expression for the perturbation of the expectation value of the CFT stress-energy tensor (\ref{A-eq:exp-calT_lower-by_K}). 
Since in our background, the extrinsic curvature is vanishing $\bar{K}_{\mu\nu} = 0$, we can ignore terms of 
$O(K^2)$. Then, we can write perturbation of eq.~(\ref{A-eq:exp-calT_lower-by_K}) as
\begin{align}
  & \delta\pExp{\calT_{\mu\nu}}
  = \lim_{z \to 0} \frac{1}{ 8 \pi G_{d+1}\, L }\,
  \Bigg\{ - L\, \left( \frac{L\, \Omega}{d - 2}\, \partial_z + 1 \right)\,
    \frac{ \delta ({\tilde K}_{\mu\nu})
      - \bar{g}_{\mu\nu}\, \bar{g}^{\rho\sigma} \delta({\tilde K}_{\rho\sigma}) }
         {\Omega^{d-1}}
  + \cdots
  \Bigg\}
  ~,
\label{A-eq:exp-calT_lower-by_K-pert}
\end{align}
where only terms which are relevant in $d < 4$ are expressed explicitly and $\cdots$ 
denotes those in $d \ge 4$, and where $\delta ({\tilde K}_{\mu\nu}) = - \partial_z h_{\mu\nu}/2$. 

The perturbation of the expectation value~(\ref{A-eq:exp-calT_lower-by_K-pert}) is written, 
by using the bulk Einstein equations~$\delta E_{MN}=0$, as
\begin{align}
  & \delta\pExp{\calT_{\mu\nu}}
  = \lim_{z \to 0} \frac{1}{ 16 \pi G_{d+1}\, L }\, \Bigg[\,
  - \left( \frac{L\, \Omega}{d - 2}\, \partial_z + 1 \right)
  \frac{ L\, \partial_z \big( \ell^2\, \bar{P}_{\mu\nu}\, \Phi \big) }
       { \Omega^{d-1} }
  + \frac{ 2\, L\, \partial_{z} \bar{D}_{(\mu} h_{T\, \nu)}^{(1)} }
         { \Omega^{d-1} }
\nonumber \\
  &\hspace*{3.0truecm}
  + \left( \frac{L\, \Omega}{d - 2}\, \partial_z + 1 \right)
    \frac{ L\, \partial_z h_{T\, \mu\nu}^{(2)} }{ \Omega^{d-1} }
  + \cdots
  \, \Bigg]
  ~, 
\label{A-eq:exp-calT_lower-by_K-pert-2-pre}
\end{align}
which is by itself gauge invariant due to the fact that $\pExp{ \bar{\cal T}_{\mu\nu}} = 0$ for our background.  
From eqs.~(\ref{A-eq:del_semiC_Einstein_eqn-LHS}) and (\ref{A-eq:exp-calT_lower-by_K-pert-2-pre}), we 
find that the perturbations of the semiclassical Einstein equation (\ref{eq:EOM-holo_semiC-Einstein_eqn-bdy}) 
is written as
\begin{align}
  & \lim_{z \to 0} \left\{
   \frac{ (d - 1)\, (d - 2) }{d}\, \bar{g}_{\mu\nu}\,
    \left( \bar{D}^2 - \frac{d}{\ell^2} \right)\, \Phi
  - \frac{d - 2}{d}\, \bar{P}_{\mu\nu}\, \Phi
  - \left( \bar{D}^2 + \frac{2}{\ell^2} \right)\, h_{T\, \mu\nu}^{(2)}
  \right\}
\nonumber \\
  &= \lim_{z \to 0} \frac{ G_{d} }{ G_{d+1}\, L }\, \Bigg[\,
  - \left( \frac{L\, \Omega}{d - 2}\, \partial_z + 1 \right)
  \frac{ L\, \partial_z \big( \ell^2\, \bar{P}_{\mu\nu}\, \Phi \big) }
       { \Omega^{d-1} }
  + \frac{ 2\, L\, \partial_{z} \bar{D}_{(\mu} h_{T\, \nu)}^{(1)} }
         { \Omega^{d-1} }
\nonumber \\
  &\hspace*{3.0truecm}
  + \left( \frac{L\, \Omega}{d - 2}\, \partial_z + 1 \right)
    \frac{ L\, \partial_z h_{T\, \mu\nu}^{(2)} }{ \Omega^{d-1} }
  + \cdots
  \, \Bigg]
  ~.
\label{A-eq:holo_semiC-Einstein_eq-pert-decomp-pre}
\end{align}
%



\begin{thebibliography}{99}  


\bibitem{Maldacena:1997re}
J.~M.~Maldacena,
``The Large N limit of superconformal field theories and supergravity,''
Adv. Theor. Math. Phys. \textbf{2}, 231-252 (1998)
[arXiv:hep-th/9711200 [hep-th]].
%
\bibitem{Gubser:1998bc}
S.~S.~Gubser, I.~R.~Klebanov and A.~M.~Polyakov,
``Gauge theory correlators from noncritical string theory,''
Phys. Lett. B \textbf{428}, 105-114 (1998)
[arXiv:hep-th/9802109 [hep-th]].
%
\bibitem{Witten:1998qj}
E.~Witten,
``Anti-de Sitter space and holography,''
Adv. Theor. Math. Phys. \textbf{2}, 253-291 (1998)
[arXiv:hep-th/9802150 [hep-th]].
%
\bibitem{Witten:2001ua}
E.~Witten,
``Multitrace operators, boundary conditions, and AdS / CFT correspondence,''
[arXiv:hep-th/0112258 [hep-th]].
%
\bibitem{Compere:2008us}
G.~Compere and D.~Marolf,
``Setting the boundary free in AdS/CFT,''
Class. Quant. Grav. \textbf{25}, 195014 (2008)
[arXiv:0805.1902 [hep-th]].
%
\bibitem{Ecker:2021cvz}
C.~Ecker, W.~van der Schee, D.~Mateos and J.~Casalderrey-Solana,
``Holographic evolution with dynamical boundary gravity,''
JHEP \textbf{03}, 137 (2022)
[arXiv:2109.10355 [hep-th]].

%
\bibitem{Natsuume:2022kic}
M.~Natsuume and T.~Okamura,
``Holographic Meissner effect,''
Phys. Rev. D \textbf{106}, no.8, 086005 (2022)
[arXiv:2207.07182 [hep-th]].

%
\bibitem{Ahn:2022azl}
Y.~Ahn, M.~Baggioli, K.~B.~Huh, H.~S.~Jeong, K.~Y.~Kim and Y.~W.~Sun,
``Holography and magnetohydrodynamics with dynamical gauge fields,''
JHEP \textbf{02}, 012 (2023)
doi:10.1007/JHEP02(2023)012
[arXiv:2211.01760 [hep-th]].
%
\bibitem{Ecker:2018ucc}
C.~Ecker, A.~Mukhopadhyay, F.~Preis, A.~Rebhan and A.~Soloviev,
``Time evolution of a toy semiholographic glasma,''
JHEP \textbf{08}, 074 (2018)
doi:10.1007/JHEP08(2018)074
[arXiv:1806.01850 [hep-th]].
%
\bibitem{Maeda:2010br}
K.~Maeda, M.~Natsuume and T.~Okamura,
``On two pieces of folklore in the AdS/CFT duality,''
Phys. Rev. D \textbf{82}, 046002 (2010)
[arXiv:1005.2431 [hep-th]].
%

\bibitem{Domenech:2010nf}
O.~Domenech, M.~Montull, A.~Pomarol, A.~Salvio and P.~J.~Silva,
``Emergent Gauge Fields in Holographic Superconductors,''
JHEP \textbf{08}, 033 (2010)
[arXiv:1005.1776 [hep-th]].
%
\bibitem{Montull:2009fe}
M.~Montull, A.~Pomarol and P.~J.~Silva,
Phys. Rev. Lett. \textbf{103}, 091601 (2009)
doi:10.1103/PhysRevLett.103.091601
[arXiv:0906.2396 [hep-th]].


\bibitem{Skenderis:2002wp}
K.~Skenderis,
``Lecture notes on holographic renormalization,''
Class. Quant. Grav. \textbf{19}, 5849-5876 (2002)
[arXiv:hep-th/0209067 [hep-th]].
%
\bibitem{Kodama:1984ziu}
H.~Kodama and M.~Sasaki,
``Cosmological Perturbation Theory,''
Prog. Theor. Phys. Suppl. \textbf{78}, 1-166 (1984).
%
\bibitem{Ishibashi:2004wx}
A.~Ishibashi and R.~M.~Wald,
``Dynamics in nonglobally hyperbolic static space-times. 3. Anti-de Sitter space-time,''
Class. Quant. Grav. \textbf{21}, 2981-3014 (2004)
[arXiv:hep-th/0402184 [hep-th]].
%
\bibitem{Banados:1992wn}
M.~Banados, C.~Teitelboim and J.~Zanelli,
``The Black hole in three-dimensional space-time,''
Phys. Rev. Lett. \textbf{69}, 1849-1851 (1992)
[arXiv:hep-th/9204099 [hep-th]].
%
\bibitem{deHaro:2000vlm}
S.~de Haro, S.~N.~Solodukhin and K.~Skenderis,
``Holographic reconstruction of space-time and renormalization in the AdS / CFT correspondence,''
Commun. Math. Phys. \textbf{217}, 595-622 (2001)
[arXiv:hep-th/0002230 [hep-th]].
%
\bibitem{Breitenlohner:1982jf}
P.~Breitenlohner and D.~Z.~Freedman,
``Stability in Gauged Extended Supergravity,''
Annals Phys. \textbf{144}, 249 (1982).
%
\bibitem{Breitenlohner:1982bm}
P.~Breitenlohner and D.~Z.~Freedman,
``Positive Energy in anti-De Sitter Backgrounds and Gauged Extended Supergravity,''
Phys. Lett. B \textbf{115}, 197-201 (1982).
%
\bibitem{Martinez:1996gn}
C.~Martinez and J.~Zanelli,
``Conformally dressed black hole in (2+1)-dimensions,''
Phys. Rev. D \textbf{54}, 3830-3833 (1996)
[arXiv:gr-qc/9604021 [gr-qc]].
%
\bibitem{Liu:2008ds}
L.~h.~Liu and B.~Wang,
``Stability of BTZ black strings,''
Phys. Rev. D \textbf{78}, 064001 (2008)
[arXiv:0803.0455 [hep-th]].
%
\bibitem{Marolf:2019wkz}
D.~Marolf and J.~E.~Santos,
``Phases of Holographic Hawking Radiation on spatially compact spacetimes,''
JHEP \textbf{10}, 250 (2019)
[arXiv:1906.07681 [hep-th]].





%
\bibitem{Steif:1993zv}
A.~R.~Steif,
``The Quantum stress tensor in the three-dimensional black hole,''
Phys. Rev. D \textbf{49}, 585-589 (1994)
[arXiv:gr-qc/9308032 [gr-qc]].
%
\bibitem{Shiraishi:1993nu}
K.~Shiraishi and T.~Maki,
``Vacuum polarization around a three-dimensional black hole,''
Class. Quant. Grav. \textbf{11}, 695-700 (1994)
[arXiv:1505.03958 [gr-qc]].
%
\bibitem{Shiraishi:1993qnr}
K.~Shiraishi and T.~Maki,
``Quantum fluctuation of stress tensor and black holes in three dimensions,''
Phys. Rev. D \textbf{49}, 5286-5294 (1994)
[arXiv:1804.07872 [gr-qc]].
%
\bibitem{Lifschytz:1993eb}
G.~Lifschytz and M.~Ortiz,
``Scalar field quantization on the (2+1)-dimensional black hole background,''
Phys. Rev. D \textbf{49}, 1929-1943 (1994)
[arXiv:gr-qc/9310008 [gr-qc]].
%
\bibitem{Hubeny:2009rc}
V.~E.~Hubeny, D.~Marolf and M.~Rangamani,
``Hawking radiation from AdS black holes,''
Class. Quant. Grav. \textbf{27}, 095018 (2010)
[arXiv:0911.4144 [hep-th]].
%
\bibitem{Hubeny:2009ru}
V.~E.~Hubeny, D.~Marolf and M.~Rangamani,
``Hawking radiation in large N strongly-coupled field theories,''
Class. Quant. Grav. \textbf{27}, 095015 (2010)
[arXiv:0908.2270 [hep-th]].
%
\bibitem{Gregory:1993vy}
R.~Gregory and R.~Laflamme,
``Black strings and p-branes are unstable,''
Phys. Rev. Lett. \textbf{70}, 2837-2840 (1993)
[arXiv:hep-th/9301052 [hep-th]].
%
\bibitem{Horowitz:2016ezu}
G.~T.~Horowitz, J.~E.~Santos and B.~Way,
``Evidence for an Electrifying Violation of Cosmic Censorship,''
Class. Quant. Grav. \textbf{33}, no.19, 195007 (2016)
[arXiv:1604.06465 [hep-th]].
%
\bibitem{Horowitz:2019dym}
G.~T.~Horowitz and D.~Wang,
``Gravitational Corner Conditions in Holography,''
JHEP \textbf{01}, 155 (2020)
[arXiv:1909.11703 [hep-th]].
%
\bibitem{Emparan:2021ewh}
R.~Emparan, D.~Licht, R.~Suzuki, M.~Toma\v{s}evi\'c and B.~Way,
``Black tsunamis and naked singularities in AdS,''
JHEP \textbf{02}, 090 (2022)
[arXiv:2112.07967 [hep-th]].
%
\bibitem{IshibashiMaedaOkamura}
A.~Ishibashi, K.~Maeda, and T.~Okamura, work in progress.  
%
\bibitem{Callan:1992rs}
C.~G.~Callan, Jr., S.~B.~Giddings, J.~A.~Harvey and A.~Strominger,
``Evanescent black holes,''
Phys. Rev. D \textbf{45}, no.4, R1005 (1992)
[arXiv:hep-th/9111056 [hep-th]].
%
\bibitem{Graham:2007va}
N.~Graham and K.~D.~Olum,
``Achronal averaged null energy condition,''
Phys. Rev. D \textbf{76}, 064001 (2007)
[arXiv:0705.3193 [gr-qc]].
%
\bibitem{Urban:2009yt}
D.~Urban and K.~D.~Olum,
``Averaged null energy condition violation in a conformally flat spacetime,''
Phys. Rev. D \textbf{81}, 024039 (2010)
[arXiv:0910.5925 [gr-qc]].

%
\bibitem{Ishibashi:2019nby}
A.~Ishibashi, K.~Maeda and E.~Mefford,
``Achronal averaged null energy condition, weak cosmic censorship, and AdS/CFT duality,''
Phys. Rev. D \textbf{100}, no.6, 066008 (2019)
[arXiv:1903.11806 [hep-th]].

%
\bibitem{Takayanagi:2011zk}
T.~Takayanagi,
``Holographic Dual of BCFT,''
Phys. Rev. Lett. \textbf{107}, 101602 (2011)
[arXiv:1105.5165 [hep-th]].

\bibitem{Izumi:2022opi}
K.~Izumi, T.~Shiromizu, K.~Suzuki, T.~Takayanagi and N.~Tanahashi,
``Brane dynamics of holographic BCFTs,''
JHEP \textbf{10}, 050 (2022)
[arXiv:2205.15500 [hep-th]]. 

\bibitem{Shiromizu:1999wj}
T.~Shiromizu, K.~i.~Maeda and M.~Sasaki,
``The Einstein equation on the 3-brane world,''
Phys. Rev. D \textbf{62}, 024012 (2000)
[arXiv:gr-qc/9910076 [gr-qc]].

%
\bibitem{Balasubramanian:1999re}
V.~Balasubramanian and P.~Kraus,
``A Stress tensor for Anti-de Sitter gravity,''
Commun. Math. Phys. \textbf{208}, 413-428 (1999)
[arXiv:hep-th/9902121 [hep-th]].



\end{thebibliography}
\end{document}